\documentclass[11pt,a4paper]{article}
\usepackage{graphicx}
\usepackage{amsmath}
\usepackage{amssymb}
\usepackage{epsfig}
\usepackage{times}
\usepackage{a4}
\usepackage[hang,stable]{footmisc}
\begin{document}

\title{Einstein's impact on the physics of the\\ twentieth century}
\author{
Domenico Giulini\\
Institute of Physics\\
University of Freiburg, Germany\\
giulini@physik.uni-freiburg.de
\and
Norbert Straumann\\
Institute for Theoretical Physics\\
University of Z\"urich, Switzerland\\
norbert.straumann@freesurf.ch}

\bigskip
\date{}
\maketitle

\begin{abstract}
\noindent
Starting with Einstein's famous papers of 1905, we review some
of the ensuing developments and their impact on present-day
physics. We attempt to cover topics that are of interest to
historians and philosophers of science as well as to physicists.
\\[0.5ex]
This paper will appear in ``2005: The Centenary of  Einstein's 
Annus Mirabilis'', the special March 2006 issue of 
\emph{Studies in History and Philosophy of Modern Physics}.
\\[0.5ex]
\end{abstract}

\newpage
\tableofcontents
\newpage

\section{Introduction}
The assignment we were given for this article was to describe the impact 
of Einstein's work on 20th-century physics. This formulation of our task 
is somewhat problematic given that a sizable fraction of 20th-century 
physics \emph{is} Einstein's work and most of the rest is more or less
directly connected to it. Hence Einstein's impact definitely
cannot be treated perturbatively. In fact, it would have been
much easier to write about those developments of 20th-century
physics that were \emph{not} connected to the work of Einstein.
But who would want to read or write that?

Einstein's major, enduring contributions to physics were made during
the first quarter of the 20th century. They can roughly be divided
into four main branches: (1)~statistical physics, (2)~early quantum
theory of light and matter, (3)~Special Relativity, and
(4)~General Relativity (theory of spacetime and gravitation).
Our article is structured accordingly, in that we will write about
each branch in turn. We regret not being able to include
material on present-day attempts to reconcile General Relativity
with Quantum Field Theory, but that would have added
another 20 pages or so to an already fairly lengthy article.

Some topics we write about seem (to us) mandatory, others are chosen
according to personal prejudices and/or predilections. Sometimes much more
could have been said, whereas in other places less detail would
have sufficed to give a first impression. For several reasons we
decided against keeping the discussions at a constant technical
level. In some cases we put more emphasis on the historical context, 
in others we chose to display some technical details. As far as the 
latter are concerned, we feel that it is important not just to recount 
the greatness of Einstein's thoughts, but also to put some flesh on 
these thoughts to see this greatness taking on a definite shape. 
In any case, we wanted to avoid letting the discussion degenerate 
into a sterile succession of ``statements of affairs''. 
In addition, we hope to address physicists with interest in the 
history and philosophy of their science as well as historians and 
philosophers of science with an interest in physics proper. 
This has called for various 
compromises. We hope to have found a readable and enjoyable balance, 
being well aware of Einstein's dictum: ``Wer es unternimmt, auf dem 
Gebiet der Wahrheit und der Erkenntnis als Autorit\"at aufzutreten, 
scheitert am Gel\"achter der G\"otter'' 
(Einstein, 1977, p.\,106).\footnote{``He who endeavors to present 
himself as an authority in matters of truth and cognition, will be 
wrecked by the laughter of the gods''. The original german text 
first appeared in (Einstein, 1952).}

\section{Einstein and statistical physics}
\subsection{A brief survey}
When Einstein's great papers of 1905 appeared in print, he was not a
newcomer to the \textit{Annalen der Physik}, in which he published most
of his early work. Of crucial importance for his further research were
three early papers on the foundations of statistical mechanics, in which he
tried to fill what he considered to be a gap in the mechanical foundations
of thermodynamics. When Einstein wrote his three papers he
was not familiar with the work of Gibbs and only partially with that of
Boltzmann. Einstein's papers, like Gibbs's 
\emph{Elementary Principles of Statistical Mechanics} 
of 1902, form a bridge between Boltzmann's work and the modern approach 
to statistical mechanics. In particular, Einstein independently 
formulated the distinction between the microcanonical and canonical 
ensembles and derived the equilibrium distribution for the canonical 
ensemble from the microcanonical distribution. Of special importance 
for his later research was the derivation of the energy-fluctuation 
formula for the canonical ensemble.

Einstein's profound insight into the nature and size of fluctuations
played a decisive role for his most revolutionary contribution to physics:
the light-quantum hypothesis. Indeed, Einstein extracted the light-quantum
postulate from a statistical-mechanical analogy between radiation in the
Wien regime\footnote{The `Wien regime' corresponds to high frequency
and/or low temperature, such that $h\nu\gg kT$, where $h$ and $k$ are
Planck's and Boltzmann's constants respectively.} and a classical ideal
gas of material particles. In this consideration Boltzmann's principle,
relating entropy and probability of macroscopic states, played a key
role. Later Einstein extended these considerations to an analysis of
energy and momentum fluctuations in the radiation field. For the latter
he also drew on ideas and methods he had developed in the course
of his work on Brownian motion, another beautiful application of
fluctuation theory. This definitively established the reality of atoms
and molecules, and, more generally, provided strong support for the
molecular-kinetic theory of thermodynamics.

Fluctuations also played a prominent role in Einstein's beautiful work
on critical opalescence. Many years later he applied this magic wand once
more to gases of identical particles, satisfying the Bose-Einstein
statistics. With this work in 1924 he extended the particle-wave duality
for photons to massive particles. It is well-known that Schr\"{o}dinger
was strongly influenced by this profound insight (see below).

\subsection{Foundations of statistical mechanics}
Already as a student Einstein was very interested in thermodynamics and
kinetic theory, and he intensively studied some of Boltzmann's work.
As he wrote on September 13, 1900
to Mileva Maric:
\begin{quote}
``The Boltzmann is absolutely magnificent. I'm almost
finished with it. He's a masterful writer. I am firmly convinced of
the correctness of the principles of the theory, i.e., I am convinced
that in the case of gases, we are really dealing with discrete mass
points of definite finite size which move according to certain
conditions. Boltzmann quite correctly emphasizes that the hypothetical
forces between molecules are not essential components of the theory,
as the whole energy is essentially kinetic in character. This is a
step forward in the dynamic explanation of physical phenomena.''
(CPAE, Vol.\,1, Doc.\,75; translation from Renn and Schulmann, 1992, 32)
\end{quote}
For further details on this incubation period, we refer to (CPAE, Vol.\,2,
editorial note, p.\,41).

The first of Einstein's three papers on the foundations of statistical
mechanics was submitted to the \textit{Annalen} in June 1902. One can only
be astonished about the self-assurance with which the 23-year-old
approaches the fundamental problems. His aim is clearly described in the
opening section:
\begin{quote}
``Great as the achievements of the kinetic theory of heat have
been in the domain of gas theory, the science of mechanics has not yet been
able to produce an adequate foundation for the general theory of heat,
for one has not yet succeeded in deriving the laws of thermal equilibrium
and the second law of thermodynamics using only the equations of mechanics
and the probability calculus, though Maxwell's and Boltzmann's theories came
close to this goal. The purpose of the following considerations is
to close this gap. At the same time, they will yield an extension of the
second law that is of importance for the application of thermodynamics.
They will also yield the mathematical expression for entropy from the
standpoint of mechanics.''
(CPAE, Vol.\,2, Doc.\,3, p.\,57)\footnote{We refer to 
\emph{The Collected Papers of Albert Einstein}
(CPAE) for all papers that have meanwhile appeared in this edition. Translations are taken from the companion volumes to the documentary editions.
}
\end{quote}
This is not the place to describe the detailed content of the three papers
(for this we refer again to the editorial note in CPAE, Vol.\,2 mentioned 
above). The third one begins with a brief polished summary of the two 
preceding ones, including several improvements. Then Einstein proceeds 
to a discussion of the ``general significance of the constant $k$'', 
by deriving the energy-fluctuation formula in the canonical ensemble.
He comments:
\begin{quote}
``Thus the absolute constant $k$ determines the thermal
stability of the system. The relationship just found is interesting
because it no longer contains any quantity reminiscent of the assumption
on which the theory is based.''
(CPAE, Vol.\,2, Doc.\,5, p.\,105)
\end{quote}
In the final section of the paper Einstein applies his fluctuation formula
to black-body radiation, a theme that would soon lead him to his
light-quantum hypothesis.

\subsection{Applications of the classical theory}
In his dissertation \emph{A new determination of molecular dimensions}, 
the second of the five papers of 1905, Einstein derived a new formula for
the diffusion constant $D$ of suspended microscopic particles.
\footnote{The main body of the paper is devoted to the derivation of a
relation between the coefficients of viscosity of a liquid with and
without suspended particles. Einstein applied this relation, together
with the diffusion formula, to the case of sugar being dissolved in
water. Using empirical data he got (after eliminating a calculational
error) an excellent value of the Avogadro number and an estimate of the
size of sugar molecules. For its wide range of applications Einstein's 
dissertation was by far the most cited of all of his papers around time 
when the Einstein biography by Pais (1982) appeared. It probably still 
is. For a recent detailed discussion of Einstein's dissertation, see 
Straumann (2005).} This formula is obtained on the basis of thermal 
and dynamical equilibrium conditions, making use of van't Hoff's law 
for the osmotic pressure and Stokes' law for the mobility of a particle. 
The result---obtained almost simultaneously by Sutherland---reads
\begin{equation}
\label{eq:E-ThesisD}
D=\frac{kT}{6\pi\eta a},
\end{equation}
where $\eta$ is the viscosity of the fluid and $a$ the radius of the
particles (assumed to be spherical).

\subsubsection*{Brownian motion}
This formula soon came to play an important role in Einstein's work
on Brownian motion. In this celebrated paper he first gives a
statistical mechanical derivation of the osmotic pressure, and then
repeats his earlier derivation of (\ref{eq:E-ThesisD}). In the short
novel part of the paper he considers the diffusion alternatively as
the result of a highly irregular random motion, caused by the
bombardment of an enormously large number of molecules. On the basis
of some idealizing assumptions, he shows that the random walks of
the suspended particles can be described by a Gaussian process,
``which was to be expected'' (CPAE, Vol.\,2, Doc.\,16, p.\,234).
Moreover, the width of the probability distribution for the position
of a particle is determined by the diffusion constant. Therefore,
the one-dimensional variance of the position is given by the famous
formula
\begin{equation}
\label{eq:E-BMotionVariance}
\left\langle(\Delta x)^2\right\rangle=2Dt=\frac{kT}{3\pi\eta_0a}~t.
\end{equation}

All this is so well-known that no further explanations are necessary.
It may, however, be appropriate to recall the following sentences of the
introductory part of Einstein's paper, which clearly express what he
considered to be important.
\begin{quote}
``If it is really possible to observe the motion to be discussed
here, along with the laws it is expected to obey, then classical
thermodynamics can no longer be viewed as strictly valid even for
microscopically distinguishable spaces, and an exact determination of the
real size of atoms becomes possible. Conversely, if the prediction of
this motion were to be proven wrong, this fact would provide a weighty
argument against the molecular-kinetic conception of heat.''
(CPAE, Vol.\,2, Doc.\,16, p.\,224)
\end{quote}

\subsubsection*{Critical opalescence}
A letter of Einstein to his collaborator Jacob Laub from August 27, 1910
(CPAE, Vol.\,5, Doc.\,224) shows his enthusiasm about his work
on critical opalescence, yet another application of the theory of
statistical fluctuations. This was Einstein's last contribution to
\emph{classical} statistical mechanics, and the corresponding measurements
were soon carried out.

Since about 1874 it was known that the scattering and attenuation of light
passing through gas becomes very large near the critical point.\footnote{
The point at which the partial derivative of the pressure with respect
to the volume at constant temperature vanishes, i.e.,
$(\partial p/\partial V)_T=0$.} In 1908 Marian von Smoluchowski pointed
out that this phenomenon is the result of density fluctuations of the
medium, but he did not derive a quantitative formula for the scattering
or extinction coefficient. Einstein set out to close this gap 
(CPAE, Vol.\,3, Doc.\,9).

Before he does, Einstein gives a lengthy introduction to
the theory of statistical fluctuations based on Boltzmann's principle.
He then applies the general theory to density fluctuations of fluids and
mixtures of fluids. This opening section is a  major
and influential contribution to statistical thermodynamics.

In the fourth section Einstein begins with the electrodynamic part of the
problem and derives the well-known formula for the scattering coefficient,
which has long become standard text-book material.
If the refraction index $n$ is close to 1, this coefficient reduces to
\begin{equation}
 \alpha(\omega)=\frac{1}{6\pi}\left(\frac{\omega}{c}\right)^4(n^2-1)^2
\frac{kT}{-V(\partial p/\partial V)_T},
 \end{equation}
 where $\omega$ is the angular frequency of the light. With this 
formula Einstein had found a quantitative relationship between 
Rayleigh scattering and critical opalescence.

At the critical point this expression diverges, because the correlation
length for the density fluctuations diverges. As was first pointed out
by Ornstein and Zernicke, Einstein's implicit assumption of statistical
independence in separated volume elements is then no longer valid.
In this sense, Einstein's work on critical opalescence became the
starting point of several research directions of the twentieth century.

 \subsection{Post-Einstein developments}
Einstein did not have a dynamical theory of Brownian motion; he determined
the nature of the motion on the basis of some assumptions. Another
derivation was later given by Langevin, who separated the force on a
suspended particle into ordered and disordered parts. Through this work he
became the founder of the theory of stochastic differential equations.
His approach was the starting point of the work of Ornstein and Uhlenbeck,
which we shall briefly discuss below. Before doing this, however, we want to
point out that Einstein's heuristic considerations, which have been
criticized by many people (including Einstein himself), are tantamount
to assumption (iii) of the following theorem.

\medskip
\noindent
 \textbf{Theorem}.
\emph{
Let $X_t~(0\leq t<\infty)$ be a stochastic process, satisfying the
 properties:
\begin{itemize}
\item[(i)]
Independence: Each increment $X_{t+\Delta t}-X_t$ is independent of
 $\{X_\tau,~\tau\leq t\}$.\\[-4.0ex]
\item[(ii)]
Stationarity: The distribution of $X_{t+\Delta t}-X_t$ does not
depend on $t$.\\[-4.0ex]
\item[(iii)]
Continuity: If $P$ denotes the probability measure belonging to the
stochastic process, then
\begin{equation}
\label{eq:StochProssIII}
\lim_{\Delta t\downarrow 0}\frac{P(\{|X_{t+\Delta t}
-X_t|\geq\delta\})}{\Delta t}=0, \quad\text{for all}\ \delta>0\,.
\end{equation}\\[-4.0ex]
\item[(iv)]
\begin{equation}
\label{eq:StochProssIV}
X_{t=0}=0.
\end{equation}
\end{itemize}
Then $X_t$ has a normal distribution with $\langle X_t\rangle=0$ and
$\langle X^2_t\rangle=\sigma^2 t$, where $\sigma$ is a numerical
constant.}

\medskip
\noindent
For a proof, see Ch.\,12 in (Breimann, 1968); see also Theorem\,5.5
in Nelson (1967).

Einstein's theory of Brownian motion is highly idealized, since for
example the velocity of a particle is not defined. Langevin's
approach, perfected by Ornstein and Uhlenbeck (Uhlenbeck, 1930), is
closer to Newtonian particle mechanics  and is thus truly dynamical.
In practice, for `ordinary' Brownian motion, the predictions of the
two theories are numerically indistinguishable.

In the Ornstein-Uhlenbeck theory the velocity process $V_t$ is described
in terms of the stochastic differential equation (Langevin equation)
\begin{equation}
\label{eq:StochProssV}
\dot{V}_t=-\alpha V_t+\sigma\xi_t\,,
\end{equation}
where $\xi_t$ denotes `white noise'. (The exact meaning of this equation
is described in every book on stochastic differential equations.)

Let us state a few important results that can be derived from the
basic equation (\ref{eq:StochProssV}).
\begin{itemize}
\item[a)]
The distribution of $V_t$ converges for large $t$ to a Gaussian
distribution with mean zero and variance $\sigma^2/2\alpha$.
Because of the equipartition theorem of statistical mechanics it is,
therefore, natural to set
$\frac{1}{2}m(\sigma^2/2\alpha)=\frac{1}{2}kT$ (where $m$ is the mass of
the particle).
The dissipation $\alpha$ thus induces a fluctuation
\begin{equation}
\label{eq:OrnUhlFluct}
\sigma^2=\frac{2\alpha}{m}kT.
\end{equation}
\item[b)]
The distributions of the positions $X_t$ converge for large $t$ to
those of the Gaussian process
\begin{equation}
\label{eq:OrnUhlGauss}
\tilde{B}_t=X_0+\sqrt{2D}B_t\,,
\end{equation}
where $B_t$ is the Brownian (Wiener) process with variance 1,
$X_0$ the initial position of the particle, and
\begin{equation}
\label{eq:OrnUhlD}
 D=\frac{\sigma^2}{2\alpha}=\frac{kT}{m\alpha}.
 \end{equation}
The distribution function of $X_t$ is thus,
\begin{equation}
\label{eq:OrnUhlDist}
p_t(x)=\frac{1}{\sqrt{4\pi Dt}}e^{-x^2/4Dt}\,,
\end{equation}
and hence satisfies the diffusion equation
\begin{equation}
\label{eq:DiffEq}
\partial_tp_t-D\partial_x^2p_t=0.
\end{equation}
 Therefore, $D$ is the diffusion constant. According to equation
(\ref{eq:OrnUhlD}) it is given by the Einstein value
(\ref{eq:E-ThesisD}), if we also use Stokes' law for $\alpha$.
\end{itemize}

The theory of stochastic differential equations has expanded into a
huge field of stochastic analysis, with rich applications in physics,
engineering, and mathematical finance. In quantum physics (generalized)
stochastic processes have become very important through Feynman-Kac path
integral representations. We briefly recall a simple example of
such a formula.

Consider on $L^2(\mathbb{R}^n)$ the Schr\"{o}dinger operator
\begin{equation}
\label{eq:SchrOp}
H=-\frac{1}{2}\triangle+V\,.
\end{equation}
Under certain conditions for the potential $V$, the operator is
self-adjoint and the following Feynman-Kac formula holds for each
$t>0$ and $\psi\in L^2$:
\begin{equation}
\label{eq:FeynKac}
\left(e^{-tH}\psi\right)(x)=\left\langle\exp\left(-\int_0^t
 V(x+B_s)~ds\right)\psi(x+B_t)\right\rangle\,,
 \end{equation}
almost everywhere in $x$. The expectation value on the right-hand side
is taken with the probability measure belonging to the Brownian process
$B_t$. Such representations have many applications (see, e.g.,
Simon, 1979).

In modern quantum field theory, path (functional) integral representations 
play a crucial role. For gauge theories they are indispensable.
A general remarkable fact, first pointed out by Feynman, is that the
Euclidean formulation of quantum field theory in terms of functional
integrals establishes a close connection with classical~(!) statistical
mechanics (models of magnetism). All this has by now become standard
text-book material (see, e.g., Roepsdorff, 1996).

\section{Einstein's contributions to quantum theory}

\subsection{Einstein's first paper from 1905}
We begin by briefly reviewing the line of thought of the March paper
(CPAE, Vol.\,2, Doc.\,14) about which Res Jost said in 1979:
``Without this paper the development of physics in our century is
unthinkable" (Jost, 1995, p.\,79). In the first section Einstein
emphasizes that classical physics inevitably leads to a nonsensical
energy distribution for black-body radiation, but that the spectral
distribution, $\rho(T,\nu)$, must approximately be correct for large
wavelengths and radiation densities (classical
regime).\footnote{This is, to our knowledge, the first proposal of a
`correspondence argument', which is of great heuristic power, as we
will see.} Applying the equipartition theorem for a system of
resonators (harmonic oscillators) in thermal equilibrium, he
independently found what is now known as the Rayleigh-Jeans law:
$\rho(\nu,T)=(8\pi\nu^2/c^3)kT$. Einstein stresses that this law
``not only fails to agree with experience (...), but is out of
question'' (CPAE, Vol.\,2, Doc.\,14, p.\,154) because it implies a
diverging total energy density (ultraviolet catastrophe). In the
second section he then states that the Planck formula, ``which has
been sufficient to account for all observations made so far''
(ibid., p.\,154) agrees with the classically derived formula in the
mentioned limiting domain for the following value of Avogadro's
number
\begin{equation}
N_A=6.17\times 10^{23}\,.
\end{equation}
This relation was already found by Planck, albeit not via a
correspondence argument. Planck relied on the strict validity of his
formula and the assumptions used in its derivation. Einstein's
correspondence argument now showed ``that Planck's determination of
the elementary quanta is to some extent independent of his theory of
black-body radiation'' (ibid., p.\,155). Indeed, Einstein understood
from first principles exactly what he did. A similar correspondence
argument was used by him more than ten years later in his famous
derivation of Planck's formula (more about this later). Einstein
concludes these considerations with the following words:
\begin{quote}
``The greater the energy density and the wavelength of
the radiation, the more useful the theoretical principles we have
been using prove to be; however, these principles fail completely
in the case of small wavelengths and small radiation densities.''
(CPAE, Vol.\,2, Doc.\,14, p.\,155)
\end{quote}

Einstein now begins to analyze what can be learned about the structure of
radiation from the empirical behavior in the Wien regime, i.e., from
Wien's radiation formula for the spectral energy-density
\begin{equation}
\label{eq:LQH-Wien}
\rho(T,\nu)=\frac{8\pi\nu^2}{c^3}h\nu e^{-h\nu/kT}\,.
\end{equation}
Let $E_V(T,\nu)$ be the energy of radiation contained in the
volume $V$ and within the frequency interval $[\nu\,,\,\nu+\Delta\nu]$
($\Delta\nu$ small), i.e.,
\begin{equation}
\label{eq:LQH-Energy}
E_V(T,\nu)=\rho(T,\nu)\,V\,\Delta\nu\,.
\end{equation}
and, correspondingly, $S_V(T,\nu)=\sigma(T,\nu)\,V\,\Delta\nu$
for the entropy. Thermodynamics implies
\begin{equation}
\label{eq:LQH-Step1}
\frac{\partial\sigma}{\partial\rho}=\frac{1}{T}\,.
\end{equation}
Solving (\ref{eq:LQH-Wien}) for $1/T$ and inserting this into
(\ref{eq:LQH-Step1}) gives
\begin{equation}
\label{eq:LQH-Step2}
\frac{\partial\sigma}{\partial\rho}
=-\frac{k}{h\nu}\ln\left[\frac{\rho}{8\pi
h\nu^3/c^3}\right]\,.
\end{equation}
Integration yields
\begin{equation}
\label{eq:LQH-Step3}
S_V=-k\frac{E_V}{h\nu}\left\{
\ln\left[\frac{E_V}{V\Delta\nu\,8\pi h\nu^3/c^3}\right]-1\right\}\,.
\end{equation}
In his first paper on this subject, Einstein focused his attention on
the volume dependence of the entropy of the radiation as given by this
expression. Fixing the amount of energy, $E=E_V,$ one obtains
\begin{equation}
\label{eq:LQH-Step4}
S_V-S_{V_0}=k\frac{E}{h\nu}\ln\left(\frac{V}{V_0}\right)=
k\ln\left(\frac{V}{V_0}\right)^{E/h\nu}\,.
\end{equation}

So far only thermodynamics has been used. Now Einstein introduces what 
he calls Boltzmann's principle, which was already of central
importance in his papers on statistical mechanics. According to Boltzmann,
the entropy $S$ of a system is connected  with the number of
possibilities $W$, by which a macroscopic state can microscopically
be realized, through the relation
\begin{equation}
\label{eq:LQH-BoltzmanPrinzip}
S=k\ln W\,.
\end{equation}
In a separate section Einstein recalls this fundamental relation between
entropy and ``statistical probability'' (Einstein's terminology)
before applying it to an ideal gas of $N$ particles in volumes $V$
and $V_0$, respectively. For the relative probability of the
two situations one has
\begin{equation}
\label{eq:LQH-BoltzmanAppl}
W=\left(\frac{V}{V_0}\right)^N\,,
\end{equation}
and hence for the entropies
\begin{equation}
\label{eq:LQH-Step5}
S(V,T)-S(V_0,T)=kN\,\ln\left(\frac{V}{V_0}\right)\,.
\end{equation}
For the relative entropies (\ref{eq:LQH-Step4}) of the radiation field,
Boltzmann's principle (\ref{eq:LQH-BoltzmanPrinzip}) now gives
\begin{equation}
\label{eq:LQH-Step6}
W=\left(\frac{V}{V_0}\right)^{ E/h\nu}\,.
\end{equation}
From the striking similarity between (\ref{eq:LQH-BoltzmanAppl}) and
(\ref{eq:LQH-Step6}) Einstein concludes:
\begin{quote}
``Monochromatic radiation of low density (within the range
of Wien's radiation formula) behaves thermodynamically as if it
consisted of mutually independent energy quanta of magnitude
$\frac{R\beta}{N}\nu$.''
(CPAE, Vol.\,2, Doc.\,14, p.\,161)
\end{quote}
Here $R\beta/N$ corresponds to $h$. So far no revolutionary statement
has been made. The famous sentences just quoted express the result of
a statistical-mechanical analysis.

\subsubsection*{Light quantum hypothesis}
Einstein's bold step consists in a statement about the quantum
properties of the free electromagnetic field that was not accepted
for a long time by anybody else. He formulates his heuristic principle
as follows (where we replaced his $R\beta/N$ by $h$):
\begin{quote}
``If, with regard to the dependence of its entropy on volume, a
monochromatic radiation (of sufficient low density) behaves like a
discontinuous medium consisting of energy quanta of magnitude
$h\nu$, then it seems reasonable to investigate
whether the laws of generation and conversion of light are so
constituted as if light consisted of such energy quanta.''
(CPAE, Vol.\,2, Doc.\,14, p.\,143-144)
\end{quote}

In the final two sections, Einstein applies this hypothesis first to
an explanation of Stokes' rule for photoluminescence and then turns to
the photoelectric effect. One should be aware that in those days only
some qualitative properties of this phenomenon were known. Therefore,
Einstein's well-known linear relation between the maximum kinetic energy
of the photoelectrons ($E_{\rm max}$) and the frequency of the incident
radiation,
\begin{equation}
\label{eq:LQH-PhotoEff}
E_{\rm max}=h\nu-P\,,
\end{equation}
was a true prediction. Here $P$ is the work-function of the metal
emitting the electrons, which depends on the material in question but
not on the frequency of the incident light. It took almost ten years
until this was experimentally confirmed by Millikan, who then used
it to give a first precision measurement of $h$ (slope of the straight
line given by (\ref{eq:LQH-PhotoEff}) in the $\nu$-$E_{\rm max}$ plane)
at the $0.5$ percent level (Millikan 1916). Strange though understandable,
not even Millikan,\footnote{Others who strongly opposed Einstein's idea, or
at least openly stated disbelief, included Planck (compare
footnote\,\ref{foot:Planck})), Sommerfeld, von\,Laue, Lorentz and Bohr.
As late as 1922, in his Nobel Lecture, Bohr (1922, p.\,14) stated that
``In spite of its heuristic value, however, the hypothesis
of light quanta, which is quite irreconcilable with so-called
interference phenomena, is not able to throw light on the nature of
radiation.'' Bohr's critical attitude culminated in his famous joint
paper of Bohr, Kramers, and Slater (1924); see, e.g.,  Section\,11d in Pais
(1991) for more background information on this fascinating episode.}
who spent 10 years on the brilliant experimental verification of
its consequence (\ref{eq:LQH-PhotoEff}), could believe in the
fundamental correctness of Einstein's hypothesis. In his comprehensive
paper on the determination of $h$, Millikan first commented on the
light-quantum hypothesis:
\begin{quote}
``This hypothesis may well be called reckless, first because an
electromagnetic disturbance which remains localized in space
seems a violation of the very conception of an electromagnetic
disturbance, and second because it flies in the face of the
thoroughly established facts of interference.''
(Millikan, 1916, p.\,355)
\end{quote}
And after reporting on his successful experimental verification
of Einstein's equation (\ref{eq:LQH-PhotoEff}) and the associated
determination of $h$, Millikan  concludes:
\begin{quote}
``Despite the apparently complete success of the Einstein
equation, the physical theory of which it was designed to be the
symbolic expression is found so untenable that Einstein himself,
I believe, no longer holds to it.''
(Millikan, 1916, p.\,384)
\end{quote}
It should be stressed that Einstein's bold light quantum hypothesis was
very far from Planck's conception. Planck neither envisaged a quantization
of the free radiation field, nor did he, as is often stated, quantize
the energy of a material oscillator per se. What he was actually doing
in his decisive calculation of the entropy of a harmonic oscillator
was to assume that the \emph{total} energy of a large number of
oscillators is made up of \emph{finite} energy elements of equal
magnitude $h\nu$. He did not propose that the energies of single material
oscillators are physically quantized.\footnote{\label{foot:Planck}
In 1911 Planck even formulated a `new radiation hypothesis', in which
quantization only applies to the process of light emission but not to
that of light absorption (Planck 1911). Planck's explicitly stated
motivation for this was to avoid an effective quantization of oscillator
energies as a \emph{result} of quantization of all interaction energies.
It is amusing to note that this new hypothesis led Planck to a
modification of his radiation law, which consisted in the addition of
the temperature-independent term $h\nu/2$ to the energy of each
oscillator, thus corresponding to the oscillator's energy at zero
temperature. This seems to be the first appearance of what soon
became known as `zero-point energy'.} Rather, the energy elements
$h\nu$ were introduced as a formal counting device that could at
the end of the calculation not be set to zero, for, otherwise, the
entropy would diverge. It was Einstein in 1906 who interpreted
Planck's result as follows (again writing $h$ for $R\beta/N)$:
\begin{quote}
``Hence, we must view the following proposition as the
basis underlying Planck's theory of radiation: The energy of an
elementary resonator can only assume values that are integral
multiples of $h\nu$; by emission and absorption, the energy of
a resonator changes by jumps of integral multiples of $h\nu$.''
(CPAE, Vol.\,2, Doc.\,34, p.\,353)
\end{quote}
\subsection{Energy and momentum fluctuations of the radiation field}
In his paper ``On the present status of the radiation problem'' of
1909 (CPAE, Vol.\,2, Doc.\,56), Einstein returned to the considerations
discussed above, but extended his statistical analysis to the entire
Planck distribution. First, he considers the energy fluctuations,
and re-derives the general fluctuation formula he had already found
in the third of his statistical-mechanics articles. This implies for
the variance of $E_V$ in (\ref{eq:LQH-Energy}):
\begin{equation}
\label{eq:E-VarianceGeneral}
\left\langle(E_V-\langle E_V\rangle)^2\right\rangle
= kT^2\frac{\partial\langle E_V\rangle}{\partial T}
=kT^2V\Delta\nu\frac{\partial\rho}{\partial T}\,.
\end{equation}
For the Planck distribution this gives
\begin{equation}
\label{eq:E-VariancePlanck}
\left\langle(E_V-\langle E_V\rangle)^2\right\rangle
=\left(h\nu\rho+\frac{c^3}{8\pi\nu^2}\rho^2\right)V\Delta\nu\,.
\end{equation}
Einstein shows that the second term  within the parentheses of this 
most remarkable formula,
which dominates in the Rayleigh-Jeans regime, can be understood with
the help of the classical wave theory as due to interference
between partial waves. The first term, dominating in the Wien
regime, is thus in obvious contradiction to classical electrodynamics.
It can, however, be interpreted by analogy to the fluctuations of the
number of molecules in ideal gases, and thus represents a particle
aspect of the radiation in the quantum domain.

Einstein confirms this particle-wave duality, at this time a genuine
theoretical conundrum, by considering momentum fluctuations.
For this he considers the Brownian motion of a mirror that perfectly
reflects radiation in a small frequency interval, but transmits radiation of
all other frequencies. About the final result he writes:
\begin{quote}
``The close connection between this relation and the one derived
in the last section for the energy fluctuation is immediately obvious,
and exactly analogous considerations can be applied to it.
Again, according to the current theory, the expression would be reduced
to the second term (fluctuations due to interference). If the first term
alone were present, the fluctuations of the radiation pressure could be
completely explained by the assumption that the radiation consists of
independently moving, not too extended complexes of energy $h\nu$.''
(CPAE, Vol.\,2, Doc.\,56, p.\,547)
\end{quote}

Einstein also discussed these issues in his famous Salzburg lecture
(CPAE Vol.\,2, Doc.\,60) at the 81st Meeting of German Scientists and
Physicians in 1909. Pauli (1949) once said that this report can be
regarded as a turning point in the development of theoretical physics.
In this lecture, Einstein treated the theory of relativity and quantum 
theory and pointed out important interconnections between his work on the
quantum hypothesis, on relativity, on Brownian motion, and statistical
mechanics. Already in the introductory section he says prophetically:
\begin{quote}
``It is therefore my opinion  that the next stage in the
development of theoretical physics will bring us a theory of light
that can be understood as a kind of fusion of the wave and emission
theories of light.''
(CPAE, Vol.\,2, Doc.\,60, p.\,564-565)
\end{quote}
We now know that it took almost twenty years until this was achieved
by Dirac in his quantum theory of radiation.

\subsubsection*{Specific heat of solids}
In 1907 Einstein used his understanding of black-body radiation
to develop a theory for the specific heat of solids (CPAE Vol.\,2, Doc.\,38).
He starts by showing that Planck's radiation law can be derived within
statistical mechanics by restricting the state sum of the oscillators
to quantized energies, and obtains for the average energy of an
oscillator the expression $h\nu/(e^{h\nu/kT}-1)$. An interesting
methodological aspect of his first paper on this subject is that
Einstein for the first time works with the canonical ensemble. He
repeatedly came back to the subject, in particular at the Solvay
Congress in 1911, when measurements by Nernst were available.
Shortly afterwards, Born and Karman and independently Debye
developed the theory that has become standard.

\subsection{Derivation of the Planck distribution}
A peak in Einstein's endeavor to extract as much information as possible about
the nature of radiation from the Planck distribution is his paper
``On the Quantum Theory of Radiation'' of 1916
(CPAE, Vol.\,6, Doc.\,38). In the first part he gives a derivation
of Planck's formula which has become part of many textbooks on
quantum theory. Einstein was very pleased by this derivation, about
which he wrote on August 11, 1916 to Besso: ``An amazingly simple
derivation of Planck's formula, I should like to say \emph{the}
derivation'' (CPAE, Vol.\,8, Doc.\,250). In this derivation
he added the hitherto unknown process of
induced emission,\footnote{Einstein's derivation shows that without
assuming a non-zero probability for induced emission one would
necessarily arrive at Wien's instead of Planck's radiation law.}
to the familiar processes of spontaneous emission and induced
absorption. For each pair of energy levels he described the
statistical laws for these processes by three coefficients
(the famous $A$- and $B$-coefficients) and established two
relations between these coefficients on the basis of his earlier
correspondence argument in the classical Rayleigh-Jeans limit
and Wien's displacement law. In addition, the latter implies
that the energy difference $\varepsilon_n-\varepsilon_m$ between
two internal energy states of the atoms in equilibrium with thermal
radiation has to satisfy Bohr's frequency condition:
$\varepsilon_n-\varepsilon_m=h\nu_{nm}$. In Dirac's 1927 radiation
theory these results follow ---without any correspondence
arguments---from first principles.

In the second part of his fundamental paper, Einstein discusses the
exchange of momentum between atoms and radiation by making
use of the theory of Brownian motion. Using a truly beautiful argument
he shows that in every elementary process of radiation, and in particular
in spontaneous emission, an amount $h\nu/c$ of momentum is emitted
in a random direction and that the atomic system suffers a
corresponding recoil in the opposite direction. This recoil was
first experimentally confirmed in 1933 by showing that a long and
narrow beam of excited sodium atoms widens up after spontaneous
emissions have taken place (Frisch, 1933).
Einstein's paper ends with the following remarkable statement
concerning the role of ``chance'' in his description of the
radiation processes by statistical laws, to which Pauli (1949)
drew special attention:
\begin{quote}
``The weakness of the theory lies, on the one hand, in the
fact that it does not bring us any closer to a merger with the
undulatory theory, and, on the other hand, in the fact that it
leaves the time and direction of elementary processes to `chance';
in spite of this I harbor full confidence in the trustworthiness
of the path entered upon.''
(CPAE, Vol.\,6, Doc.\,38, p.\,396)
\end{quote}

\subsection{Bose-Einstein statistics for degenerate material gases}
The last major contributions of Einstein to quantum theory were
stimulated by de\,Broglie's suggestion that material particles also have
a wave aspect, and Bose's derivation of Planck's formula, which
only made  use of the picture of light as particles, albeit particles satisfying a new statistics
on account of their indistinguishability. Einstein (1924, 1925a, 1925b) applied
Bose's statistics for photons to degenerate gases of identical
massive particles. With this `Bose-Einstein statistics', he obtained
a new law, to become known as the Bose-Einstein distribution. As with
radiation, Einstein considered fluctuations in these gases and
found both particle-like and wave-like aspects. This time the
wave property was the novel feature that was recognized by Einstein
to be necessary.

In the course of this work on quantum gases, Einstein discovered the
condensation of such gases at low temperatures. (Although Bose made
no contributions to this, one nowadays speaks of Bose-Einstein
condensation.) Needless to say that this subject has become enormously
topical in recent years.

In his papers on wave mechanics, Schr\"{o}dinger acknowledged the
influence of Einstein's gas theory, which from today's perspective
appear to be his last great constructive contribution to physics
proper. In the article in which Schr\"{o}dinger establishes the
connection of matrix and wave mechanics, he remarks in a footnote:
``My theory was inspired by L.\,de\,Broglie and by brief but
infinitely far-seeing remarks of A.\,Einstein [1925a, p.\,9 ff.]"
(Schr\"{o}dinger, 1926, p.\,735).

It is well-known that Einstein considered the `new' quantum mechanics
to be unsatisfactory until the end of his life. In his
autobiographical notes, for example, he writes:
\begin{quote}
``I believe, however, that this theory offers no useful point
of departure for future developments. This is the point at which my
expectation departs most widely from that of contemporary
physicists.''
(Einstein, 1979, p.\,83)
\end{quote}

\subsection{Light quanta after 1925}
In his contribution to one of the foundational papers on matrix 
mechanics (Born \& Jordan 1925),
Pascual Jordan made it clear  that the quantum-interpretation of
physical observables must apply to the electromagnetic field as well.
He elaborated on this in
the extended final section of the \textit{Dreim\"{a}nnerarbeit} by
Born, Heisenberg and Jordan. In particular, Jordan derived Einstein's
fluctuation formula (\ref{eq:E-VariancePlanck}) from a
description of the cavity radiation as an infinite set of uncoupled
harmonic oscillators, quantized according to the rules of matrix
mechanics.\footnote{This was inspired by earlier work of Ehrenfest~(1906)
and Debye~(1910).} With this and later investigations, partly in
collaboration with other authors (Klein, Wigner, Pauli), Jordan is
not only one of the creators of quantum mechanics, but also one of
the founding fathers of quantum field theory.\footnote{For biographical
notes we refer to Sec.\,1.2 of (Schweber, 1994).}

After Jordan Dirac was the first  to address, in the fall of 1926,
the quantum-theoretic description of the electromagnetic field.
In this seminal work he treated for the first time the quantized
electromagnetic field in interaction with atomic matter described by
non-relativistic wave mechanics. Treating the coupled system in first
order perturbation theory, he obtained directly---without the use of
correspondence arguments---Einstein's rules for emission and absorption
of light. As Gregor Wentzel wrote in an article on the
early history of quantum field theory in the memorial volume
for Wolfgang Pauli,
\begin{quote}
``Today, the novelty and boldness of Dirac's approach to the
radiation problem may be hard to appreciate. During the preceding decade
it had become a tradition to think of Bohr's correspondence principle
as the supreme guide in such questions, and, indeed, the efforts to
formulate this principle in a quantitative fashion had led to the
essential ideas preparing the eventual discovery of matrix mechanics
by Heisenberg. A new aspect of the problem appeared when it became
possible, by quantum mechanical perturbation theory to treat atomic
transitions induced by given external wave fields, e.g.,  the photoelectric
effect. The transitions so calculated could be interpreted as being
caused by absorptive processes, but the ``reaction on the field'',
namely the disappearance of a photon, was not described by the theory,
nor was there any  possibility, in this framework, of understanding the
process of spontaneous emission. Here, the correspondence principle
still seemed indispensable, a rather foreign element (a ``magic wand''
as Sommerfeld called it) in this otherwise very coherent theory.
At this point, Dirac's explanation in terms of the $q$ matrix came
as a revelation. Known results were re-derived, but in a completely
unified way. The new theory stimulated further thinking about
application of quantum mechanics to electromagnetic and other
fields.''
(Wentzel, 1960, p.\,49)
\end{quote}

In Dirac's theory the dual particle/wave aspects of radiation are
described in a coherent, logically consistent manner. The shortcomings
of the theory, however, were immediately pointed out by Ehrenfest and others.
Since the interaction terms contain the vector potential at the
position of the point-like electron, the theory would lead to
infinities in higher-order perturbation theory. In particular, the
self-energy of a free or bound electron turned out to be infinite.
Because of these divergence difficulties most theorists working
on problems in quantum electrodynamics problems in
those early days had little faith
in the theory. In Sec.\,\ref{sec:QFT} we shall take up this subject again
and sketch the further developments of relativistic quantum field
theory.

\subsection{Einstein and the interpretation of quantum mechanics}
\label{sec:QMinterp}
The new generation of young physicists who participated in the
tumultuous three-year period from January 1925 to January 1928 deplored
Einstein's negative judgement of quantum mechanics. In the
article on Einstein's contributions to quantum mechanics cited above,
Pauli expressed the disappointment of his contemporaries:
\begin{quote}
``The writer belongs to those physicists who believe that the
new epistemological situation underlying quantum mechanics is
satisfactory, both from the standpoint of physics and from the
broader knowledge in general. He regrets that Einstein seems to
have a different opinion on this situation (...).''
(Pauli, 1949, p.\,149)
\end{quote}
When the Einstein-Podolsky-Rosen (EPR) paper (Einstein
\emph{et\,al}.~1935) appeared, Pauli's immediate reaction
in a letter to  Heisenberg of June 15th was quite furious:
\begin{quote}
``Einstein once again has expressed himself publicly on quantum
mechanics, namely in the issue of \emph{Physical Review} of 
May 15th (in cooperation with Podolsky and Rosen -- not a good 
company, by the way). As is well known, this is a catastrophe each 
time when it happens.'' (Pauli, 1985--99, Vol.\,2, Doc.\,412, p.\,402)
\end{quote}
From our present vantage point this judgment is clearly too harsh,
but it shows the attitude of the `younger generation' towards
Einstein's concerns. In fact, Pauli understood (even if he did not accept)
Einstein's point much better than many others, as his intervention
in the Born-Einstein debate on Quantum Mechanics shows (Born 2005;
Pauli to Born, March 31, 1954).
Whatever one's attitude on this issue is, it is certainly true that
the EPR argumentation has engendered an uninterrupted discussion up to
this day. The most influential of John Bell's papers on the
foundations of quantum mechanics bears the title ``On the
Einstein-Podolsky-Rosen paradox'' (Bell 1964). In this publication
Bell presents what has come to be called ``Bell's Theorem'', which
(roughly) asserts that \emph{no hidden-variable theory that 
satisfies a certain locality condition can produce all 
predictions of quantum mechanics}. This signals the importance of 
EPR's paper in focusing on a pair of well-separated particles that 
have been properly prepared to ensure strict correlations between 
some of the observable quantities associated with them. Bell's 
analysis and later refinements (Bell, 1987) showed clearly that 
the behavior of entangled states is explicable only in the language 
of quantum mechanics.

This point has also been the subject of the very interesting, but much
less known work of Kochen \& Specker (1967), with the title
``The Problem of Hidden Variables in Quantum Mechanics''. Loosely
speaking, Kochen and Specker show that quantum mechanics
\emph{cannot} be embedded in a classical stochastic theory, provided
two very desirable conditions are assumed to be satisfied.
The first condition (KS1) is that the quantum-mechanical distributions
are reproduced by the embedding of the quantum description into a
classical stochastic theory. (The precise definition of this concept is
given in the cited paper.) The authors first show that hidden variables
in this sense can always be introduced if there are no other requirements.
(This is not difficult to prove.) The second condition (KS2)
states that a function $u(A)$
of self-adjoint operators $A$ representing quantum-mechanical observables
has to be represented in the classical description
by the very same function $u$ of the image $f_A$ of $A$, where $f$
is the embedding that maps the operator $A$ to the classical
observable $f_A$ on `phase space'. Formally, (KS2) states
that for all $A$
\begin{equation}
\label{eq:KS2}
f_{u(A)}=u\left(f_A\right).
\end{equation}

The main result of Kochen and Specker states that if the dimension of
the Hilbert space of quantum mechanical states is larger than 2, an
embedding satisfying (KS1) and (KS2) is `in general'
\emph{not possible}.

There are many highly relevant examples---even of low dimensions with
only a finite number of states and observables---where this
impossibility holds.

The original proof of Kochen and Specker is very ingenious, but quite
difficult. In the meantime several authors have given much simpler
proofs (e.g., Straumann, 2002).

We find the result of Kochen and Specker entirely satisfactory in the
sense that it clearly demonstrates that there is no way back to
classical reality. Einstein's view that quantum mechanics is a kind
of glorified statistical mechanics that ignores some hidden microscopic
degrees of freedom, can thus not be maintained without giving up
locality or (KS2). It would be interesting to know his reaction to
these developments triggered by the EPR paper.

Entanglement is not limited to questions of principle. It has already
been employed in quantum communication systems, and entanglement
underlies all proposals of quantum computation.

\section{Special Relativity as a symmetry principle}

\subsection{Historical origin and conceptual meaning}
\label{sec:HistOrigin}
The principle of relativity goes back at least to Galileo.
The idea that mechanical experiments cannot reveal an overall uniform
and inertial (rectilinear) motion became known as the `Galilean Principle
of Relativity'. In Newtonian mechanics it is expressed mathematically
by the invariance of its equations of motion under the Galilean group.
This mathematical statement has two interpretations, whose physical
connotations differ in a subtle way. The first interpretation,
called the `passive' one, is that of a mere change of reference frames
while keeping the system under study fixed. In the second interpretation,
called the `active' one, one keeps the reference system fixed while changing
the state of motion of the system under study. If the physical world just
consisted of these two objects, the reference system and the system
under study, these two interpretations would be equivalent, since both
amount to stating a \emph{relative} change in the state of motion and
there is nothing more to state. However, this is not the situation usually encountered in physics. Typically, one has a system $S$ to be
studied, a reference frame $F$ (which can be thought of as a physical 
system in its own right), and the rest $R$ of the physical universe, 
parts of which may at times interact with $S$ but which can otherwise be 
neglected. In the passive interpretation we only change the frame $F$, 
that is, we change the relative state of motion between $F$ and the 
totality of other systems, here denoted by $S+R$. In the active 
interpretation we only act on $S$, in which case the cut is between 
$S$ and $F+R$.

So even if dynamically silent, the presence of $R$ is important for
the interpretation of symmetries. This is because a symmetry
connects \emph{physically distinguishable} states, thereby mapping
solutions of the equations of motion to other, distinguishably different
solutions. In the language of Hamiltonian mechanics this means that
the Hamiltonian function that generates the motion is invariant
under the symmetry operation, but other observables need not be.
This is precisely the difference between a physical symmetry and
a gauge transformation. Unfortunately this difference is sometimes
blurred by speaking of ``gauge symmetries''.

After the establishment of the principle of relativity in mechanics,
the natural question to ask was whether non-mechanical phenomena could 
reveal preferred states of inertial motion. Such a preference was 
strongly suggested by various `ether' theories of light  and other 
electromagnetic phenomena during the 19th century. In fact, Newton 
already expressed his firm belief in some sort of force-mediating 
`ether'. In a famous letter to Robert Bentley,
Newton wrote in 1692:
\begin{quote}
``That gravity should be innate inherent \& essential to matter
so yt one body may act upon another at a distance through a vacuum wthout
the mediation of any thing else by \& through wch their action of force
may be conveyed from one to another is to me so great an absurdity that
I believe no man who has in philosophical matters any competent faculty
of thinking can ever fall into it.''
(Newton, 1961, p.\,254)
\end{quote}

All optical and electromagnetic experiments, however, failed to show any 
trace of an ether rest-frame. This was hard to reconcile with Maxwell's
equations, which predicted an invariable speed $c$ for electromagnetic
waves in matter free space, and which were therefore thought to hold only 
in the ether's rest frame. The solution to this problem was first given 
by Lorentz (1904) and Poincar\'e (1906). They found that instead of being 
Galilean invariant\footnote{It does not seem to be widely appreciated that 
a precise statement of Galilean non-invariance needs to invoke restrictive
assumptions concerning the type of action, such as locality.
It is instructive and amusing to note that there exists a non-local
implementation of the Galilean group which makes it a symmetry group
of Maxwell's equations; see, e.g., Sec.\,5.9 in (Fushchich
\emph{et\,al.}, 1993).} Maxwell's equations are invariant under the 
Lorentz group. Einstein independently derived this result in his 1905 
paper on Special Relativity (henceforth abbreviated SR), but, unlike 
Lorentz and Poincar\'e, gave a direct physical meaning to the Lorentz 
transformations in terms of measurements of lengths and 
times.\footnote{See (Damour, 2005) for a lucid recent account on 
Poincar\'e's contribution to SR.} One may say that Einstein established 
them on a \emph{kinematical} rather than \emph{dynamical} basis, 
though one should add here that this distinction is only defined relative 
to the assumption that the ``rods'' and ``clocks'' entering the kinematical 
considerations eventually obey dynamical laws compatible with Lorentz 
invariance. If this is granted, the FitzGerald-Lorentz contraction, for 
example, can be understood kinematically (i.e., as a result of a 
fundamental symmetry that is postulated to be realized by all fundamental 
matter-equations) rather than dynamically (i.e., as consequence of a 
complicated dynamical interaction between the measuring rod and the ether). 
Note that these two viewpoints are not mutually exclusive.\footnote{In
this respect Pauli wrote in his 1921 review article on Relativity:
``The contraction of a measuring rod is not an elementary but a very
complicated process. It would not take place except for the covariance with
respect to the Lorentz group of the basic equations of electron theory,
as well as those laws, as yet unknown to us, which determine the cohesion
of the electron itself" (Pauli, 1958, p.\,15).} But the shift in 
emphasis establishes a symmetry principle with potentially far 
superior heuristic power.

In summary it seems fair to say that in 1905 SR seemed palpably close
after all the preliminary work done by various people. But
apparently it needed an unprejudiced newcomer to take the final step.

\subsection{The Lorentz group}
\label{sec:LorentzGroup}
The new understanding of the Lorentz transformations as
fundamental symmetries induced a very powerful selection
principle for dynamical laws: All fundamental dynamical laws of
Nature should be Lorentz invariant.\footnote{The reader should be
aware that there is some confusion in the literature as to the
different meanings of terms like `invariant', `covariant', etc.}
By this we mean: (1)~there is an action of the Lorentz group
on state space; (2)~this action maps solution curves to solution
curves. (An alternative but equivalent definition uses observables
rather than states.) After Minkowski's seminal work, as a result
of which SR was gradually put into its modern
mathematical form, this task could be approached in a systematic
fashion.

Minkowski realized that the Lorentz group could be understood as
the automorphism group of a geometric structure on spacetime,
which is as follows: The model for spacetime is a four-dimensional
real affine\footnote{The affine structure of spacetime is usually
motivated by the law of inertia, by means of which one identifies
inertial trajectories with (a subset of) the `straight lines'
of affine geometry.} space whose underlying vector space,
$\mathbb{R}^4$, is endowed with a non-degenerate, symmetric bilinear
form $\eta$ of signature $(-,+,+,+)$.\footnote{In the present context
it is a matter of convention whether one chooses $(-,+,+,+)$ [`mostly plus']
or  $(+,-,-,-)$ [`mostly minus']. But in more exotic situations,
like for non-orientable spacetimes, the overall sign generally matters;
see, e.g., (DeWitt-Morette \& DeWitt, 1990).} In appropriate
coordinates one has $\eta_{\mu\nu}=\text{diag}(-1,1,1,1)$.
$\eta$ is called the \emph{Minkowski metric} and the affine space endowed
with it is called \emph{Minkowski space}. The homogeneous Lorentz group
is then characterized as the set of invertible linear transformations
that leave $\eta$ invariant.

The Galilean group, too, can be characterized as the automorphism
group of some geometric structure on spacetime, which is again
modelled on real four-dimensional affine space. The `geometry' now
includes an absolute simultaneity structure and a fixed euclidean
metric on the simultaneity hypersurfaces.  We stress that, at least
as far as mechanics is concerned, the usual terminology `non-relativistic'
versus `relativistic' is quite inappropriate. Newtonian mechanics is
perfectly relativistic: the principle of relativity being implemented
by the Galilean group. What distinguishes Lorentz-invariant from 
Galilean-invariant mechanics is not the validity of the relativity 
principle, but the structurally different implementations of it.

The major structural differences between the (homogeneous, proper,
orthochronous) Galilean group and the Lorentz group is, that the 
latter is simple,\footnote{A group is called \emph{simple} if it 
has no non-trivial (i.e. other than the group itself and the group 
formed by the neutral element alone) invariant subgroups. It is 
called \emph{semi-simple} if it has no non-trivial abelian invariant 
subgroups.} whereas the former is not even semi-simple due to the 
invariant abelian subgroup formed by the pure boost transformations. 
In contrast, for the Lorentz group, the set of pure boosts do not 
even form a subgroup.
This is more than just a mathematical curiosity. It implies that
the relation of `being relatively unrotated' is not transitive among
inertial reference frames. If $K'$ is boosted relative to $K$ and
$K''$ is boosted relative to $K'$, then $K''$ is boosted
\emph{as well as rotated} relative to $K$, unless the boost
velocities of $K'$ and $K''$ are collinear. A well known early
application  of this feature, which is not present in the Galilean group, was
the downward correction by 50\% of the spin-orbit coupling and
consequently of the fine-structure intervals in atomic spectra,
which was first pointed out by Thomas (1927).\footnote{A manifestly
Lorentz invariant treatment using the Dirac equation automatically
takes care of this effect.} The same effect is even more pronounced
in nuclear physics, where the strong acceleration due to the nuclear
force leads via the `Thomas correction' to a much larger spin-orbit
coupling than that due to the electromagnetic interaction, thereby
giving rise to the so-called `inverted doublets'. The non-transitivity
of the relation `being-relatively-unrotated' has more recently also
entered the discussion of large-scale astronomical reference
frames.\footnote{See, e.g.,  (Klioner \& Soffel, 1998). For a review 
of algebraic and geometric aspects of the Lorentz group see, e.g.,  
(Giulini, 2005b).}

\subsection{Far-reaching consequences}
\label{sec:}
Replacing the Galilean group with the Lorentz group requires a modification
of the dynamical laws of mechanics, since the latter is supposed
to act through dynamical symmetries. The `heuristic power' associated with
this requirement now comes to the fore (cf.~the end of
Sec.\,\ref{sec:HistOrigin}) . Consider, e.g.,  the simplest
case of a free point-particle of mass $m_0$. Classically its dynamics is
fully described by an action whose Lagrangian is just its kinetic
energy, $\frac{1}{2}m_0v^2$. A straightforward Lorentz invariant
modification, which approaches the classical law in the limit
$c\rightarrow \infty$, is given by the action
\begin{equation}
\label{eq:ActionPointParticle}
S_{\text{particle}}=-m_0c^2\int d\tau=-m_0c^2\int\sqrt{1-v^2/c^2}\,dt\,,
\end{equation}
where $d\tau=\sqrt{-\eta_{\mu\nu}dz^\mu dz^\nu}$ is the proper time
along the worldline $z^\mu(t)$ of the particle---obviously a
Lorentz-invariant quantity. From the Lagrangian
$L=-m_0c^2\sqrt{1-v^2/c^2}$, the expressions for
energy and momentum immediately follow by standard Lagrangian methods
(again we set $\gamma(v) \equiv 1/\sqrt{1-v^2/c^2}$):
\begin{alignat}{3}
\label{eq:EnergyPointParticle}
& E &&\, \equiv \,\vec v\cdot\frac{\partial L}{\partial\vec v}-L
&&\,=\,\gamma(v)\,m_0c^2\,,\\
\label{eq:MomentumPointParticle}
& \vec p &&\, \equiv \,\quad\frac{\partial L}{\partial\vec v}
&&\,=\,\gamma(v)\,m_0\vec v\,.
\end{alignat}
Together they form the \emph{momentum four-vector} $p^\mu=(E/c,\vec p)$,
which under a Lorentz transformation, given by the matrix $L^\mu_\nu$,
transforms like
\begin{equation}
\label{eq:FourMomentumLorentzTrans}
p^\mu\mathop{\longrightarrow}^L
{p'}^\mu=L^\mu_\nu p^\nu\,.
\end{equation}
Clearly, the Minkowski-square of the four-momentum
is an invariant (we write $p \equiv \vert\vec p\vert$):
\begin{equation}
\label{eq:MomentumMinkSquare1}
\eta_{\mu\nu}p^\mu p^\nu=p^2-E^2/c^2=-m_0^2c^2\,,
\end{equation}
showing that the following relation between energy and momentum
is a Lorentz covariant one:
\begin{equation}
\label{eq:MomentumMinkSquare2}
E^2=c^2\,(p^2+m_0^2c^2)\,.
\end{equation}
This equation replaces the familiar $E=p^2/2m_0$ of Newtonian
mechanics and plays a central role throughout special-relativistic
quantum (field) theory. One of its prominent features is that
$E$ enters quadratically.

These somewhat formal derivations
(no interactions have been discussed yet) can be complemented by
an analysis of elastic two-particle scattering processes, which shows that
(\ref{eq:MomentumPointParticle}) is the unique generalization
of the classical equation $\vec p=m_0\vec v$  compatible with
momentum conservation and Lorentz invariance.\footnote{See, e.g.,
(Giulini, 2005a) for a brief presentation of this argument,
which goes back to Lewis \& Tolman (1909)}
From (\ref{eq:MomentumPointParticle}) one may deduce the expression
for the kinetic energy, $E_{\text{kin}}=m_0c^2(\gamma(v)-1)$, which
is just (\ref{eq:EnergyPointParticle}), properly normalized so that
$E_{\text{kin}}=0$ for $v=0$.

The normalization of energy in (\ref{eq:EnergyPointParticle})
is not determined by general methods (which always allow for
additive constants). The last of Einstein's five papers of
1905, just about three pages long, shows that the normalization
adopted in  (\ref{eq:EnergyPointParticle}) is more than just
a convenient choice. More precisely, using (1)~the principle
of relativity, (2)~conservation of energy, (3)~the existence of
a Newtonian limit, and (4)~the transformation law for the energy
of an electromagnetic wave, as derived from the Lorentz
transformation properties of the electromagnetic field, Einstein
shows that any emission of electromagnetic radiation with
energy $\Delta E$ by a body must decrease its inertial
rest mass $m_0$ by $\Delta E/c^2$.\footnote{See Stachel \& Torretti (1982)
for a careful account of Einstein's argument, which also saves it
from an unwarranted but often repeated criticism.} He further
argues that this holds independently of the form into which the
energy extracted from the body is turned. From this he jumps to the
conclusion that \emph{all} of the inertial mass of a body is a
measure of its energy content; later this was expressed in the now
most famous formula
\begin{equation}
\label{eq:Emc2}
E=mc^2\,.
\end{equation}

The implications of this far reaching insight can hardly be
overrated. It provided the first means to estimate the
enormous magnitude of nuclear binding-energies. Today
(\ref{eq:Emc2}) is often taken as a symbolic expression for the
ambivalent `nuclear age'. But it should be stressed that
(\ref{eq:Emc2}) only allows to `weigh' binding energies.
It neither explains them nor does it explain any of the nuclear
processes, like fission or fusion, which belong to the realm
of nuclear physics proper. The weight of binding energies becomes
even dominant on sub-nuclear scales. For example, according to
Quantum Chromodynamics, the mass of a proton (made up of three
light quarks, two `up' and one `down', interacting
via gluon exchange) is almost entirely due to interaction
energies. The quark masses themselves contribute only about 2\%.

On a more fundamental level (\ref{eq:Emc2}) changed our concept
of matter radically, in that it opens up the possibility for
different forms of matter to change into each other. To be sure,
the `channels' along which these transmutations occur are
constrained by various conservation laws. But there can be no
doubt that this puts an irreversible end to the idea of naive
atomism, since everlasting and unchanging elementary objects
simply cannot exist. Rather, modern high-energy particle physics
speaks and thinks in terms of \emph{creation} and 
\emph{annihilation} processes.

\subsection{The current experimental status of SR}
\label{sec:SRTexp}
Modern particle physics would be unthinkable without
SR. Leaving aside the conceptual implications just mentioned,
it has far-reaching kinematical consequences.
For example, proton-antiproton collisions at Fermilab's Tevatron
take place at energies  of about 2\,TeV, which is 2000 times the
rest energy of the proton. In such machines there clearly is ample
opportunity for possible deviations from SR
to manifest themselves. Since these experiments, however, are not
primarily designed to test SR, the quantities
observed in them will depend in complicated ways on the
fundamental assumptions of SR. This makes
it hard to infer good quantitative upper-bounds for
violations of SR from such experiments, even if,
energetically speaking, they take place in the ``ultrarelativistic
regime''.

Experiments specifically designed to test the principle of
relativity basically probe for dynamical effects of preferred
reference frames. A good candidate for such a preferred frame is
one in which the cosmic microwave background (CMB) appears most isotropic
(i.e., without dipole anisotropy). It is called the CMB-frame.
Ever since observation of the dipole anisotropy with the Cosmic 
Background Explorer (COBE)  we know that the barycenter of our solar 
system moves relative to the CMB-frame at a speed of $370\,km/s$ 
(Kogut\,\emph{et\,al.}~1993).

Let us suppose that the CMB-frame, $K$, is such that in $K$ the
velocity of light $c=1$ (in appropriately chosen units) in all directions.
Note that this implies that clocks in $K$ are Einstein-synchronized.
The transformation formulae between $K$ (coordinatized by $(\vec x,t)$) 
and an inertial frame $K'$ (coordinatized by $(\vec x',t')$) moving
with relative velocity $\vec v\equiv\vec n v$ ($\vec n\cdot\vec n=1$) 
with respect to $K$ are then of the general
form (Mansouri \& Sexl,~1976)
\begin{equation}
\label{eq:CMBframeTrans}
\begin{split}
\vec x' &= d(v)\vec x+\vec n(\vec n\cdot\vec x)
\bigl(b(v)-d(v)\bigr)-b(v)\vec v\,t\,,\\
     t' &= a(v)t+\vec\varepsilon(\vec v)\cdot\vec x'\,.
\end{split}
\end{equation}
Here $a,b,d$ are functions of $v$ whose interpretation is easily
inferred: $a$ is the factor of time dilation (for this reason we
wrote $t'$ as function of $\vec x'$ rather than $\vec x$), and
$b$ and $d$ are the factors of longitudinal and transverse length
contraction respectively. These functions are to be determined
experimentally. The values that SR assigns to them are
\begin{equation}
\label{eq:SRValues}
a_{\text{SR}}(v)=1/b_{\text{SR}}(v)=\sqrt{1-v^2}\,,
\quad d_{\text{SR}}(v)=1\,.
\end{equation}
The vector $\vec\varepsilon$  is determined by
$a,b,d$ once the convention for clock synchronization in $K'$
is chosen. For example, for Einstein-synchronization one has
\begin{equation}
\label{eq:EinSynch}
\vec\varepsilon(\vec v)=\vec\varepsilon_{\text{E}}(\vec v) \equiv
-\,\vec v\,\frac{a(v)}{b(v)(1-v^2)}\,,
\end{equation}
leading to the familiar expression
$\vec\varepsilon_{\text{SR}}(\vec v)=-\vec v$ for SR. If we agree 
to Einstein-synchronize clocks in $K'$,
the following expression can be derived for the velocity of light
in $K'$ (Mansouri \& Sexl 1976)\footnote{The relevant formula in
this reference, (6.17), has a misprint: $d^2$ in the denominator
should be $d^{-2}$, as shown in (\ref{eq:VelLight}).}
\begin{equation}
\label{eq:VelLight}
c'(\theta,v)=
\frac{b(v)(1-v^2)}{a(v)
\sqrt{\cos^2\theta+b^2(v)d^{-2}(v)(1-v^2)\sin^2\theta}}\,,
\end{equation}
where $\theta$ is the angle between the light ray and $\vec v$
as measured in $K'$. This reduces to $c'=1$ for the values given
in (\ref{eq:SRValues}), but depends on $\theta$ in the general
case. The invariance under $\theta\rightarrow\theta+\pi$ reflects 
the Einstein-synchronization of the clocks in $K'$.

We want to stress the following conceptually very important point:
The expression for $c'(\theta,v)$ depends on the choice of clock
synchronization in $K'$. This means that if one uses it to
calculate light travel-times along open paths (i.e., paths that 
do not begin and end at the same point in space), the result will 
also depend on that choice. However, the calculated travel times will be
\emph{independent} of one's synchronization convention if the
light paths are closed in space, since in that case only a single clock
is involved. This is the case in the Michelson-Morley and
Kennedy-Thorndike experiments discussed below.

To second order in $v$ we have for $a,b,d$:
\begin{equation}
\label{eq:abcPara}
a(v)\approx 1+\alpha v^2\,,\quad
b(v)\approx 1+\beta v^2 \,,\quad
d(v)\approx 1+\delta v^2\,.
\end{equation}
Experiments checking round-trip travel times of light involve $1/c'$,
which to second order is given by:
\begin{equation}
\label{eq:InvC}
\frac{1}{c'(\theta,v)}\approx
1+(\beta-\delta-\tfrac{1}{2})v^2\sin^2\theta+
(\alpha-\beta+1)v^2\,.
\end{equation}
In SR one has $\alpha=-\beta=-1/2$
and $\delta=0$ so the expressions in parentheses vanish.

Experiments checking the $\theta$ dependence of $c(\theta,v)$
are commonly referred to as ``Michelson-Morley'' experiments; 
those checking the $v^2$ dependence as ``Kennedy-Thorndike'' experiments.
The most stringent upper-bounds  for the relative $\theta$-variation
of $c'(\theta,v)$ provided by modern measurements are of the order of
$10^{-15}$. For the $v$-variation, they are of the order $10^{-12}$. 
To translate these results into statements about the coefficients
$(\beta+\delta-\tfrac{1}{2})$ and $(\alpha-\beta+1)$ one has
to assume some value for $v$, that is, one has to make an
assumption about the value of our present velocity with respect to 
the potentially preferred frame. Since the latter is presently unanimously
stipulated to be the CMB-frame,\footnote{Though it is considered to
be unlikely, it is not impossible that the gravitational-wave
background---once it is observed---`moves' relative to the CMB
frame and therefore defines another potentially preferred frame.}
with respect to which we move at a speed of  $370\,km/s$ or
$1.23\cdot 10^{-3}$ times the speed of light, one
sets $v=1.23\cdot 10^{-3}$. With this value, the best current 
estimates (at the one-$\sigma$ level) of the upper-bound for the 
coefficients in (\ref{eq:InvC}) are (see M\"uller \emph{et\,al.}, 2003;
Wolf\,\emph{et\,al.}, 2003):
\begin{alignat}{3}
\label{eq:MMbound}
& \vert\beta-\delta-\tfrac{1}{2}\vert &&\ <\ 3.7\cdot 10^{-9}
\qquad && (\text{MM--experiment})\,,\\
\label{eq:KTbound}
& \vert\alpha-\beta+1\vert            &&\ <\ 6.9\cdot 10^{-7}
\qquad && (\text{KT--experiment})\,.
\end{alignat}
To obtain upper-bound for the three parameters, $\alpha$, $\beta$, and 
$\delta$, an independent third experiment is needed. Experiments that 
allow independent determination of the factor $\alpha$ related to time 
dilation are called ``Ives-Stilwell'' experiments. In the latest version 
one does such experiments using so-called double Doppler-spectroscopy 
(with Lasers) on ${}^7\text{Li}^+$ ions, moving at a speed of 
$19\,000\,km/s$. The best value today is (Saathoff, 2003):
\begin{equation}
\label{eq:ISbound}
\vert2\alpha+1\vert < 2.2\cdot 10^{-7}
\qquad\text{(IS--experiment)}\,.
\end{equation}
For more on the most recent experimental situation in SR, see
(Ehlers \& L\"ammerzahl, 2006)

The upper-bound on the value of $\alpha$ has additional
conceptual significance. We mentioned that Einstein-synchronization 
in $K'$ fixes $\vec\varepsilon$ to be the function given by 
(\ref{eq:EinSynch}). Now, as was probably first realized by Eddington 
(1924, p.\,11), in SR Einstein-synchronization is equivalent to 
synchronization by ``slow clock-transport''. In the more general 
setting discussed here, one can show that the value for $\vec\varepsilon$ 
corresponding to slow clock-transport is given by
\begin{equation}
\label{eq:SlowTransSynch}
\vec\varepsilon(\vec v)=\vec\varepsilon_{\text{T}}(\vec v)\, \equiv \,
\vec n\ \frac{a'(v)}{b(v)}\,,
\end{equation}
where $\vec n=\vec v/v$ and $a'$ is the derivative of $a$. Hence the 
two synchronizations agree if and only if the expressions in 
(\ref{eq:EinSynch}) and (\ref{eq:SlowTransSynch}) agree. This is 
the case when $a(v)=a_{\text{SR}}(v)$ (we obviously require that 
$a(v)=1$ for $v=0$). The upper-bound (\ref{eq:ISbound}) may therefore 
also be read as the upper-bound for possible discrepancies between 
Einstein synchronization and synchronization by slow clock-transport.

\subsection{Relativistic quantum field theory}
\label{sec:QFT}
From the very beginning, Lorentz invariance was  a guiding principle
in the development of quantum theory. Black-body radiation belongs
above all to the quantum theory of the electromagnetic field, which had
to be relativistically invariant. In this connection an important step
by Jordan and Pauli (1928) should be mentioned. These authors introduced 
time-dependent field operators for the charge-free Dirac radiation field
and determined the commutators of two field components evaluated at
different spacetime points. For the field operators $F_{\mu\nu}(x)$
these commutators can be expressed in a manifestly invariant form with the help
of the now famous invariant Jordan-Pauli distribution.
The physical meaning of these results in terms of basic uncertainty
relations in field measurements was later clarified by
Bohr \& Rosenfeld (1933).

As is well-known, Schr\"{o}dinger originally considered a Lorentz invariant  
equation, now known as the Klein-Gordon equation. Since this equation gave 
the wrong fine-structure splitting, Schr\"{o}dinger restricted himself  to
the more modest goal of a non-relativistic wave mechanics. The
decisive next step was taken by Dirac who succeeded in generalizing Pauli's
description of spin-$\frac{1}{2}$ particles to a relativistic wave equation.
Initially, Dirac's theory was considered a single-particle theory, but this
interpretation was beset with great difficulties coming from negative
energy states. These states could not consistently be eliminated and
time-dependent external fields could cause transitions from positive to
negative energy states.

\subsubsection*{Reinterpretation of Dirac's single particle theory}
Dirac's solution to this problem was his so-called `hole theory'.
The ground state then becomes stable because all negative energy states
are considered occupied so that transitions of positive energy
electrons into negative energy states are forbidden by the Pauli
Exclusion Principle. Furthermore, the vast `sea' of negative energy
particles is declared to be invisible. A `hole' in this sea was
interpreted by Dirac as a particle of positive energy and positive
charge. At first, Dirac suggested that such particles be identified
with the proton. It was soon pointed out, however, by
Oppenheimer~(1930) that this was unacceptable because it would imply 
that the hydrogen atom be very short-lived. Dirac accepted this
criticism and proposed the existence of anti-electrons:
\begin{quote}
``A hole, if there were one, would be a new kind of particle,
unknown to experimental physics, having the same mass and opposite
charge to the electron. We should not expect to find any of them in
nature, on account of their rapid rate of recombination with electrons,
but in high vacuum, they would be quite stable and amenable to
observations.''
(Dirac, 1931, p.\,61)
\end{quote}

Many of Dirac's colleagues were shocked by the audacity of his
ideas. As an example we recall Pauli's skepticism, expressed in his
famous article on wave mechanics~(1933) before the discovery of the
positron. First he point out that if there were anti-electrons there
should also be anti-protons. He then writes: ``The factual absence
of such particles then is reduced to a special initial state, in
which there is indeed only one kind of particles. This appears to be
unsatisfactory already because of the fact that the laws of nature
in this theory are symmetrical with respect to electrons and
anti-electrons'' (Pauli, 1933, p.\,246). The matter-antimatter
asymmetry is still a major problem of cosmology, about which we
shall make some remarks in Sec.\,\ref{sec:CPT}.

After Anderson's discovery of the positron, it became clear that future
work on quantum electrodynamics (QED) of spin-$\frac{1}{2}$ particles had
to be based on hole theory, or something closely related to it. Through the
work of Jordan and Wigner it became clear that the Dirac field had to be 
quantized by imposing anti-commutation relations. With the resulting elegant 
formalism it was possible to write the theory of electrons and positrons in 
a completely symmetric form under exchange of particles and anti-particles. 
In this formulation, which can be found in any modern quantum-field-theory 
textbook, the Dirac sea has no place ``except as a poetic description for 
forming the electromagnetic current'' (Wightman, 1972,\,p.\,100).

Heisenberg and Pauli (1929, 1930) were the first to attempt a general
formulation of QED as a dynamical relativistic theory of quantized fields.
With all these developments a revolution had taken place that was driven
by the problem of reconciling quantum mechanics and special relativity.
All further developments are based on these foundational pillars.
Below we shall make a few remarks about the tortuous and ongoing
history of quantum field theory. What is amazing is that  this theory, despite
all its intrinsic difficulties, makes the most precise
predictions in all of physics. What exactly lies behind this
success is still unclear.

\subsubsection*{Renormalization theory}
In the early 1930s a number of processes, such as radiative pair
creation and annihilation, were successfully computed in the Born
approximation. But in higher orders troublesome divergences
remained. Weisskopf showed that compared to the single electron theory
the most divergent terms for the self-energy cancelled, but a
logarithmic divergence remained. It was realized only after World War
II that this remaining divergence would also disappear after a mass
renormalization. A central problem early on was that of vacuum
polarization. This was studied by a number of authors. Anticipating
the idea of charge renormalization, they were able to extract correct finite
predictions for observable effects, e.g., in the energies of bound electrons.
In  even higher orders in the fine structure constant, a
fascinating phenomenon turned up: Maxwell's equations are corrected
by very small non-linear terms in the field strengths and their
derivatives, leading for instance to photon-photon scattering.
Heisenberg's subtraction procedure lead to finite expressions, as
was shown by Euler, Kockel and Heisenberg (Euler \& Kockel, 1935;
Heisenberg \& Euler, 1936)). Shortly afterwards, Weisskopf (1936)
not only simplified their calculations but also gave a thorough
discussion of the physics involved in charge renormalization.
Weisskopf related the modification of the Lagrangian of Maxwell's theory
to the change of the energy of the Dirac sea as a function of slowly varying
external electromagnetic fields. (Avoiding the old fashioned Dirac sea,
one could now interpret
this effective Lagrangian in terms of the interaction of a classical
electromagnetic field with the vacuum fluctuations of the electron
positron field.) After a charge renormalization this change is finite
and gives rise to electric and magnetic polarization vectors of the
vacuum. These investigations showed that the quantum vacuum has very
interesting properties, a subject we shall take up in
Sec.\,\ref{sec:RelCosm} in connection with the current Dark Energy
problem.

Notwithstanding these successes, most of the leading physicists were not
happy with the subtraction procedures and repeatedly expressed their
misgivings. In the late 1940s renormalization theory was developed in
a systematic manner with the help of new, manifestly Lorentz invariant
techniques. The infinities could then be sidestepped in an unambiguous
manner. The new powerful methods of Feynman, Schwinger, Tomonaga,
and Dyson made it possible to perform higher-order perturbation calculations for
QED which turned out to be in spectacular agreement with experiment. With these
developments QED became one of the most brilliant successes in the
history of physics.

Quantum field theory provides answers to some of the most profound
questions about the nature of matter. It explains why there are two
classes of particles---fermions and bosons---and how their properties
are related to their intrinsic spin (spin-statistics theorem).
The mysterious nature of indistinguishability in quantum mechanics is
understood, because identical particles are created by the same
underlying field.

QED became a model for non-Abelian gauge theories and the development
of the highly successful Standard Model of particle physics. Since the
early history of gauge theories is strongly tied to General Relativity
(henceforth abbreviated GR), we postpone further discussion both of this subject,
and of more recent developments, which include the gravitational
interaction, to Sec.\,\ref{sec:GaugeKKTh}.

\subsection{Group-theoretic background of relativistic quantum field theory}
\label{sec:GroupQFT}
Mathematically speaking, the content of SR is largely the
requirement of Lorentz invariance. A characterization of the
impact of SR on other branches of physics should therefore also
include some statements about specific properties that can be traced to
this requirement. This is particularly interesting in
Quantum Field Theory, where aspects of representation theory
become important. The representation theory as such, however, can
be discussed using classical rather than quantized fields.

For simplicity we ignore space and time reflections and consider
the group $\mathbb{R}^4\rtimes SL(2,\mathbb{C})$, which is the
double (and universal) cover of the connected component of
the inhomogeneous Lorentz group. In what follows, we will simply
refer to it as the Poincar\'e group.

The classical fields $\Psi$ under consideration are maps from spacetime
(Minkowski space) to some vector space $V$. The space $V$ carries a finite
dimensional irreducible representation $D^{(p,q)}$ of $SL(2,\mathbb{C})$.\footnote{Here
$p$ and $q$ are zero or integer multiples of $1/2$. $2p$ and $2q$
denote the numbers of `undotted' and `dotted' spinor indices respectively,
carried by the field.} With the appropriate choice of an inner
product, the infinite-dimensional linear space of such fields
carries a unitary representation of the Poincar\'e group.
The free (i.e., linear) classical field equations of
Klein-Gordon, Weyl (Neutrino equation), Dirac, Maxwell, Proca,
Rarita-Schwinger, Bargmann-Wigner and Pauli-Fierz can then collectively
be understood as projection conditions onto irreducible subspaces
(possibly including space and time reflections) in this space.

Investigations into the representation theory of the Poincar\'e group
started with a seminal paper by Wigner (1939). This was one of
the first serious mathematical papers on the representation theory of
non-compact Lie groups.\footnote{Remarkably, before sending it to
\emph{Annals of Mathematics}, Wigner submitted his paper to the
less prestigious \emph{American Journal of Mathematics}, where it
was rejected with the remark that ``this work is not interesting
for mathematics'' (see Wigner, 1993, Part A, Vol\,1, p.\,9).}
Later Mackey generalized Wigner's method to what is now known as
the theory of induced  representations. This  `Mackey Theory'
reduces to the present case if one specializes to semi-direct
products with one Abelian factor (here the translations).
A nice account of this is given, e.g., by Niederer \& O'Raifeartaigh~(1974).

Wigner's construction of irreducible representations can briefly be
described as follows: first one replaces the field $\Psi(x)$
on spacetime by its Fourier transform $\tilde\Psi(k)$ on momentum space.
An element $(a,A)\in \mathbb{R}^4\rtimes SL(2,\mathbb{C})$ acts on
$\tilde\Psi(k)$ via
\begin{equation}
\label{eq:Wigner1} \tilde\Psi(k)\mapsto 
e^{ik\cdot a}D^{(p,q)}(A)\tilde\Psi(A^{-1}k)\,.
\end{equation}
This immediately shows that irreducible subspaces must consist of fields
whose support is confined to a single group orbit in momentum space.
These orbits decompose into the following types: two families of
infinitely many orbits each, indexed by mass $m>0$ and given by the
two-sheet hyperbolas $k_0=\pm\sqrt{m^2+\vec k^2}$, the future and
past light cone, one infinite family (indexed by $\mu>0$)  of
one-sheet hyperbolas  $\vert\vec k\vert=\sqrt{\mu^2+k_0^2}$, and
finally the origin $k=0$.

The condition of having support within a group orbit, say of the first
type, translates into a differential equation for $\Psi$, which in case
of orbits of the first two types ($m>0$) is just the Klein-Gordon
equation for each component of $\Psi$:
\begin{equation}
\label{eq:WignerKG} (\Box-m^2)\Psi=0.
\end{equation}

If we restrict ourselves to functions with support on one such orbit, say $O$,
Wigner's trick consists of picking a reference point $k_*$ on $O$
and an element $A_k\in SL(2,\mathbb{C})$ for each $k$ on $O$ such that
$A_kk_*=k$. Using $A_k$, Wigner now redefines the basic field as follows:
\begin{equation}
\label{eq:Wigner2a} \tilde\Psi_W(k) \equiv D^{(p,q)}(A^{-1}_k)\tilde\Psi(k).
\end{equation}
The field $\tilde\Psi_W$ obeys a transformation law of the form of
(\ref{eq:Wigner1}), the only difference being that $D^{(p,q)}(A)$
gets replaced by
\begin{equation}
\label{eq:Wigner2b}
D^{(p,q)}(W(k,A))\,,\quad\text{where}\quad
W(k,A):=A^{-1}_{Ak}AA_k\,.
\end{equation}
What may look like a complication is, in fact, a crucial
simplification, due to the obvious fact that $W(k,A)k_*=k_*$.
One says that $W(k,A)$ lies in the subgroup
$\text{Stab}(k_*)\subset SL(2,\mathbb{C})$ of elements that fix
(`stabilize') $k_*$. One thus sees that an irreducible
representation of the Poincar\'e group is obtained by
imposing a simple projection condition on $\tilde\Psi_W$,
saying that it assumes values in a subspaces of $V$ that
is irreducible under the group $\text{Stab}(k_*)$.
If translated back to the fields $\Psi(x)$, such conditions
become the wave equations which complement
a condition such as (\ref{eq:WignerKG}) of having support
on one orbit only. Regarding the groups $\text{Stab}(k_*)$,
one has
\begin{equation}
\label{Wigner3}
\text{Stab}(k_*)=
\begin{cases}
SU(2)    &\text{for $k_*$ timelike (massive case)}\\
\bar E(2)&\text{for $k_*$ lightlike and $\ne 0$ (massless case)}\\
SL(2,\mathbb{R}) &\text{for $k_*$ spacelike (tachyonic case)}\\
SL(2,\mathbb{C}) &\text{for $k_*=0$\,.}
\end{cases}
\end{equation}

The massive cases are thus classified according to the value for
mass (picking the orbit) and spin (classifying the unitary irreducible
representation of the stabilizer subgroup, here $SU(2)$).
The massless cases are classified according
to the unitary irreducible representations of $\bar E(2)$, the double
cover of two-dimensional Euclidean motions. Here there are many more
representations than seem physically relevant. Those which represent the
`translations' in $\bar E(2)$ non-trivially are all infinite dimensional
and are usually discarded (they correspond to infinitely many `internal' 
degrees of freedom). The remaining representations of the one-parameter 
subgroup of rotations are classified by a single number, helicity, which 
is either zero or a positive-integer multiple of $1/2$. The remaining 
cases have so far not found applications with a clear physical interpretation,
though they appear in various guises in some versions of string theory. 
All non-trivial unitary irreducible representations of $SL(2,\mathbb{R})$ 
and $SL(2,\mathbb{C})$ are necessarily infinite dimensional and were 
classified by Bargmann (1947). One is thus left with the massive and 
massless cases.

The irreducible representation-spaces are Hilbert spaces
(of square integrable functions on an orbit in momentum space),
which in the physically relevant cases are denoted by $\mathcal{H}_{m,s}$
(where $s$ refers to spin for $m>0$ and to helicity for $m=0$).
In relativistic quantum field theory they serve as \emph{definition}
of `one-particle Hilbert spaces', which are used as elementary
building blocks for the total Hilbert space. This is where the dictum,
often attributed to Wigner, comes from that an `elementary particle'
\emph{is} a unitary irreducible representation of the Poincar\'e group.

In relativistic quantum field theories processes of pair creation
and annihilation are dynamically unavoidable. Hence it would be
inconsistent to limit oneself to one-particle spaces $\mathcal{H}_{m,s}$.
Particles of type $(m,s)$ should be represented by their entire
\emph{Fock space}
\begin{equation}
\label{eq:Fock}
\mathcal{F}_{m,s}=\bigoplus_{n\in\mathbb{N}}
\mathcal{H}_{m,s}^{\otimes n}\,,
\end{equation}
where $\otimes n$ either denotes the symmetrized (for $2s$ even) or
antisymmetrized (for $2s$ odd) $n$-fold tensor product. The total
Hilbert space is then the tensor product over all Fock spaces for
all particle species considered. This is the arena where scattering
states in perturbative Quantum Field Theory live.\footnote{As a
consequence of a theorem due to Rudolf Haag, it is known
that Fock space cannot be the representation space for the
fundamental equal-time commutation relations in case of
translation invariant theories of interacting fields (see,
e.g., the later (reprint) edition of Streater \& Wightman,
1963). Fock space, however, still plays a useful role for 
displaying scattering states and S-matrices.}

\subsection{The rise of supersymmetry}
\label{sec:Supersymm}
One issue that attracted much attention during the 1960s was,
whether the observed particle multiplets could be understood on the basis of
an all embracing symmetry principle that would combine the
Poincar\'e group with the internal symmetry groups displayed by
the multiplet structures. This combination should be non-trivial,
i.e., not a direct product, for otherwise the internal symmetries would
commute with the spacetime symmetries and lead to
multiplets degenerate in mass and spin (see, e.g.,
O'Raifeartaigh, 1965). Subsequently, a number of no-go theorems
appeared, which culminated in
the now most famous theorem of Coleman \& Mandula~(1967). This theorem
states that those generators of symmetries of the $S$-matrix
belonging to the Poincar\'e group necessarily commute with those
belonging to internal symmetries. The theorem is based on a series of
assumptions\footnote{The assumptions are:
(1)~there exists a non-trivial (i.e., $\not =\mathbf{1}$) S-matrix
   which depends analytically on $s$ (the squared center-of-mass
   energy) and $t$ (the squared momentum transfer);
(2)~the mass spectrum of one-particle states consists of (possibly
   infinite) isolated points with only finite degeneracies;
(3)~the generators (of the Lie algebra) of symmetries of the $S$-matrix
   contains (as a Lie-subalgebra) the Poincar\'e generators;
(4)~some technical assumptions concerning the possibility of
   writing the symmetry generators as integral operators in
   momentum space.}
involving the crucial technical condition that the $S$-matrix depends 
analytically on standard scattering parameters. What is less visible 
here is that the structure of the Poincar\'e group enters in a decisive 
way. This result would not follow for the Galilean group, as was explicitly 
pointed out by Coleman \& Mandula (1967).

One way to avoid the theorem of Coleman \& Mandula is to generalize
the notion of symmetries. An early attempt was made by Golfand \&
Likhtman (1971), who constructed what is now known as a Super-Lie
algebra, which generalizes the concept of Lie algebra (i.e.
symmetry generators obeying certain commutation relations) to
one also involving anti-commutators. In this way it became possible
for the first time to link particles of integer and half-integer spin
by a symmetry principle. It is true that supersymmetry still maintains
the degeneracy in masses and hence cannot account for the mass
differences in multiplets. But its most convincing property,
the symmetry between bosons and fermions, suggested a most elegant
resolution of the notorious ultraviolet divergences that beset
Quantum Field Theory.

It is remarkable that the idea of a cancellation of bosonic and
fermionic contributions to the vacuum energy density occurred to
Pauli. In his lectures on ``Selected Topics in Field Quantization'',
delivered in 1950-51 and still in print, he posed the question
``whether these zero-point energies
[from Bosons and Fermions]
can compensate each other''
(Pauli, 2000, p.\,33).
He tried to answer this question by writing down the formal
expression for the zero point energy density of a quantum field of spin
$j$ and mass $m_j>0$ (Pauli restricted attention to spin $0$ and
spin $1/2$, but the generalization is immediate):
\begin{equation}
\label{eq:PauliVenergy1}
4\pi^2\frac{E_j}{V}=
(-1)^{2j}(2j+1)\int dk\,k^2\sqrt{k^2+m^2}\,.
\end{equation}
Cancellation should take place for high values of $k$.
The expansion
\begin{equation}
\label{eq:PauliVenergy2}
4\int_0^Kdk\,k^2\sqrt{k^2+m^2}=
K^4+m_j^2K^2-m_j^4\log(2K/m_j)
+O(K^{-1})
\end{equation}
shows that the quartic, quadratic, and logarithmic terms must
cancel in the sum over $j$ for the limit $K\rightarrow\infty$
to exist. This implies that for $n=0,2,4$ one must have
\begin{equation}
\label{eq:PauliVenergy3}
\sum_j(-1)^{2j}(2j+1)m_j^n=0\quad \text{and}\quad
\sum_j(-1)^{2j}(2j+1)\log(m_j)=0\,.
\end{equation}
Commenting on this result, Pauli observed that
``these requirements are so extensive that it is rather
improbable that they are satisfied in reality''
(Pauli, 2000, p.\,33).

The idea of supersymmetry is that this is precisely what happens
as a consequence of the one-to-one correspondence between bosons 
and fermions. But the real world does not seem to be as simple as that.
Supersymmetry, if it exists at all, must strongly be broken in
the phase we live in. So far no supersymmetric partner of any
existing particle has been detected, even though some of them
(e.g., the neutralino) are currently suggested to be viable
candidates for the missing-mass problem in cosmology.
Future findings (or non-findings) at the Large Hadron Collider 
(LHC) will probably have a decisive impact on the future of the 
idea of supersymmetry, which---whether or not it is realized in 
Nature---is certainly very attractive.

\subsection{More on spin-statistics}
\label{sec:SpinStat}
Pauli's proof of the spin-statistics correlation is such an
impressive example for the force of abstract symmetry principles,
that we wish to recall the basic lemmas on which it rests.
We begin by replacing the proper orthochronous Lorentz
group by its double (= universal) cover $\mathrm{SL}(2{,}\mathbb{C})$ to include half-integer
spin fields.
We stress that everything that follows merely requires the
invariance under this group. No requirements concerning invariance
under space- or time reversal are needed.

Any finite-dimensional complex representation of
$\mathrm{SL}(2{,}\mathbb{C})$ is labelled by an ordered pair $(p,q)$,
where $p$ and $q$ may assume independently all non-negative integer
or half-integer values.\footnote{As before, $2p$ and $2q$ are the 
numbers of `undotted' and `dotted' spinor indices, respectively.} 
The tensor product of two such representations decomposes as follows
\begin{equation}
\label{eq:ClebschGordan}
D^{(p,q)}\otimes D^{(p',q')}=
\bigoplus_{r=\vert p-p'\vert}^{p+p'}\quad
\bigoplus_{s=\vert q-q'\vert}^{q+q'}\, D^{(r,s)}\,,
\end{equation}
where---and this is the important point in what follows---the sums
proceed in \emph{integer} steps in $r$ and $s$.
With each $D^{(p,q)}$ let us associate a `Pauli Index',
given by
\begin{equation}
\label{eq:PauliIndex}
\pi: D^{(p,q)}\rightarrow ((-1)^{2p}\,,\,(-1)^{2q})
\ \in\ \mathbb{Z}_2\times\mathbb{Z}_2\,.
\end{equation}
This association may be extended to sums of such $D^{(p,q)}$
proceeding in integer steps, simply by assigning to the sum
the Pauli Index of its terms (which are all the same).
Then we have\footnote{This may be expressed by saying that the
map $\pi$ is a homomorphism of semigroups. One semigroup consists of
direct sums of irreducible representations proceeding in integer
steps with operation $\otimes$, the other is
$\mathbb{Z}_2\times\mathbb{Z}_2$, which is actually a group.}
\begin{equation}
\label{eq:PauliIndexTensor}
\pi(D^{(p,q)}\otimes D^{(p',q')})=\pi(D^{(p,q)})\cdot\pi(D^{(p',q')})\,.
\end{equation}

According to their representations, we can associate a Pauli Index with
spinors and tensors. For example, a tensor of odd/even degree
has Pauli Index $(-,-)$/$(+,+)$. The partial derivative, $\partial$,
counts as a tensor of degree one. Now consider the most general linear
(non interacting) field equations for integer spin (here and in what
follows $\sum (\cdots)$ simply stands for ``sum of terms of the
general form $(\cdots)$''):
\begin{equation}
\label{eq:FreeFieldEq}
\begin{split}
\sum\partial_{(-,-)}\Psi_{(+,+)}&\,=\,\sum \Psi_{(-,-)}\,,\\
\sum\partial_{(-,-)}\Psi_{(-,-)}&\,=\,\sum \Psi_{(+,+)}\,.\\
\end{split}
\end{equation}
These are invariant under
\begin{equation}
\label{eq:Theta}
\Theta:
\begin{cases}
\Psi_{(+,+)}(x)\ \mapsto\  &\hspace{-5mm}\Psi_{(+,+)}(-x),\\
\Psi_{(-,-)}(x)\ \mapsto\ -&\hspace{-5mm}\Psi_{(-,-)}(-x)\,.
\end{cases}
\end{equation}
Next consider any current that is a polynomial in the fields
and their derivatives:
\begin{equation}
\label{eq:Current}
\begin{split}
J_{(-,-)}= \sum\ &\Psi_{(-,-)}
 + \Psi_{(+,+)}\Psi_{(-,-)}
 + \partial_{(-,-)}\Psi_{(+,+)} \\
 + &\Psi_{(+,+)}\partial_{(-,-)}\Psi_{(+,+)}
 + \Psi_{(-,-)}\partial_{(-,-)}\Psi_{(-,-)}
 + \quad\cdots\\
\end{split}
\end{equation}
Then one has
\begin{equation}
(\Theta J)(x)=-J(-x)\,.
\end{equation}
This shows that for any solution of the field equations with charge $Q$
for the conserved current $J$ ($Q$ being the space integral over $J^0$) there
is another solution (the $\Theta$ transformed) with charge $-Q$.
It follows that charges of conserved currents cannot be sign-definite in
any $\mathrm{SL}(2{,}\mathbb{C})$-invariant theory of non-interacting
integer spin fields. In the same fashion one shows that
conserved quantities, stemming from divergenceless symmetric
tensors of rank two, bilinear in fields, cannot be sign-definite
in any $\mathrm{SL}(2,\mathbb{C})$ invariant theory of non-interacting
half-integer spin fields. In particular, the conserved quantity in
question could be energy!

An immediate but far reaching first conclusion is
that there cannot exist a relativistic generalization of Schr\"odinger's
one- particle wave equation. For example, for integer-spin particles,
one simply cannot construct a non-negative spatial probability
distribution derived from conserved four-currents. Hence these
results for c-number fields strongly indicate the need for second quantization.

Upon second quantization the celebrated spin-statistics connection,
first proven by Fierz (1939), 
can be derived in a few lines. It says that integer spin fields
cannot be quantized using anticommutators and half-integer spin field
cannot be quantized using commutators. Here the already mentioned
Jordan-Pauli distribution plays a crucial role\footnote{The
Jordan-Pauli distribution is  uniquely
characterized (up to a constant factor) by: 
(1) it is Lorentz invariant; 
(2) it vanishes for spacelike separated arguments; 
(3) it satisfies the Klein-Gordon equation. The (anti)commutators of 
the free fields must be proportional to the Jordan-Pauli distribution, 
or to finitely many derivatives of it, either of exclusively even or of
exclusively odd order.} in the (anti)commutation relations, which ensures
causality (observables localized in spacelike separated regions
commute). Also, the crucial hypothesis of the existence of an
$\mathrm{SL}(2,\mathbb{C})$ invariant stable vacuum state is adopted.
Pauli ends his paper by saying:
``In conclusion we wish to state, that according to our opinion
the connection between spin and statistics is one of the most
important applications of the special relativity theory.''
(Pauli, 1940, p.\,722). It took almost 20 years before first attempts 
were made to generalize this result to the physically relevant case of 
interacting fields by L\"uders \& Zumino (1958).

\subsection{Existence of antimatter (CPT-theorem)}
\label{sec:CPT}
A remarkable general consequence of local relativistic quantum field
theory is the existence of antimatter, even if the theory is not
invariant under charge conjugation (C). The CPT-theorem states that
the invariance with respect to the proper Lorentz group implies the
anti-unitary CPT symmetry. ($P$ stands for space reflection and $T$ for
time reflection.) In the framework of Lagrangian field theory several
authors (Schwinger, L\"{u}ders, and others) contributed to this important
result, but the final formulation was given by Pauli~(1955), assuming,
besides locality, the normal spin-statistics connection. Soon afterwards,
Jost (1957) gave a general proof of the CPT-theorem using Wightman's
framework of quantum field theory (`axiomatic quantum field theory').
In fact, he proved a more precise result. Jost's refined form of the
theorem states that the CPT symmetry holds if and only if the following
\emph{weak locality condition} is satisfied: Consider, for simplicity,
a theory with a single neutral scalar field $\varphi(x)$. In that case, 
the vacuum expectation values of products of field operators (Wightman 
distributions) satisfy
\begin{equation}
(\Omega,\varphi(x_1)\varphi(x_2)\cdot\cdot\cdot\varphi(x_n)\Omega)=
(\Omega,\varphi(x_n)\varphi(x_{n-1})\cdot\cdot\cdot\varphi(x_1)\Omega),
\end{equation}
if the $\{x_j\}$ are pairwise spacelike: $(x_i-x_j)^2>0$ for all $i\neq j$.
The elegant proof of Jost (1957), which was the starting point for many
applications, makes crucial use of the elementary fact that the
simultaneous reflection in space and time is contained in the
identity-component of the complex Lorentz
group. (See also the classic books of Jost (1965), and of Streater \&
Wightman (1963).) The CPT-theorem has become very important, because the
electro-weak interactions are not invariant under the separate operations
$C$, $P$ and $T$. It has many applications, and so far no sign of an
experimental violation of the CPT-symmetry has been found. Because of
this deeply rooted symmetry, the observed matter-anti-matter asymmetry
in the universe is a profound problem. In spite of interesting attempts,
no satisfactory quantitative explanation has been put forward. For a
recent review, see Dine \& Kusenko (2004).

\section{On the journey to General Relativity}
\label{sec:JourneyGR}

It is often said that whereas SR was  ``in the air'' around
1905, GR would hardly be  conceivable without the penetrating thinking
of Albert Einstein. His path to GR meandered, encountered
confusing forks, and even included a major U-turn. Einstein's
own words to describe the ambivalent feelings of the searching
mind are unforgettable
\begin{quote}
``Im Lichte bereits erlangter Erkenntnisse erscheint das
gl\"ucklich Erreichte fast wie selbstverst\"andlich,
und jeder intelligente Student erfa{\ss}t es ohne zu
gro{\ss}e M\"uhe. Aber das ahnungsvolle, Jahre w\"ahrende
Suchen im Dunkeln mit seiner gespannten Sehnsucht, seiner
Abwechslung von Zuversicht und Ermattung und seinem endlichen
Durchbrechen zur Klarheit, das kennt nur, wer es selbst
erlebt hat.''
(Einstein, 1977, p.\,138)\footnote{``In the light of knowledge attained, the happy achievement seems almost a matter of course, and any intelligent student can grasp it without too much trouble. But the years of anxious searching in the dark, with their intense longing, their alternations of confidence and exhaustion and the final emergence into the light---only those who have experienced it can understand it."}
\end{quote}

This is not the place to give an account of the complex history 
that led from SR to GR (see Renn, forthcoming). But what we can do 
here is to present some selected issues from a physicist's 
perspective. We start with some early attempts to formulate a 
relativistic theory of gravity and then turn to the question how  
GR could have been discovered within the framework of Poincar\'e 
invariant field theories.

\subsection{Early attempts}
Soon after the formulation of SR Einstein began thinking about how 
to fit Newtonian gravity within that framework. Already in his 
``Jahrbuch paper'' (CPAE, Vol.\,2, Doc.\,47) he went beyond the 
framework of SR. He did not seriously consider the possibility of a 
special-relativistic theory of gravity until presented with such a 
theory by Gunnar Nordstr\"om (Norton 1992, 1993). Except for his 
attempted rebuttals of Nordstr\"om's theories no notes appear to be 
extant to document his own early attempts in this direction. But later 
recollections by Einstein make it quite easy to more or less guess the
essential steps. The following contains our (modern) interpretation
of how one might proceed along the lines of Einstein's 1933
recollections (reprinted in Einstein, 1977, english translation in 
Einstein, 1954). There he says:
\begin{quote}
``The simplest thing was, of course, to retain the Laplacian
scalar potential of gravity, and to complete the Poisson
equation in an obvious way by a term differentiated with
respect to time in such a way, that compatibility with
special relativity was achieved.''
(Einstein, 1977 p.\,135)
\end{quote}
Einstein obviously refers to replacing the Laplacian $\Delta$ by 
the d'Alembertian
\begin{equation}
\label{eq:ScalGrav1}
\Box=\Delta-\frac{\partial^2}{c^2\partial t^2}\,,
\end{equation}
 in the Poisson equation
\begin{equation}
\label{eq:ScalGrav0}
\Delta\phi=4\pi G\,\rho
\end{equation}
where $\phi$ is the gravitational potential, $\rho$ is the mass density, 
and $G$ is Newton's gravitational constant.

This turns the left-hand side of the Poisson equation into
a Lorentz-scalar. But then the source term on
the right hand side of (\ref{eq:ScalGrav0}) should also be a scalar,
which is neither true for the density of mass nor for
the density of rest-mass (rest-mass is a scalar, but its density is
not). Later Laue pointed out to Einstein that the trace
$T=T^\mu_\mu$ of the energy-momentum tensor was a natural
candidate, as Einstein e.g. acknowledges at the end of the 
``Entwurf'' paper (cf. CPAE, Vol.\,4, Doc.\,13, p.\,322). This 
leads to\footnote{\label{foot:T-Sign} Recall that $T^0_0=-T_{00}$ 
is minus the energy density, due to our convention
$\eta_{\mu\nu}=\text{diag}(-,+,+,+)$.}
\begin{equation}
\label{eq:ScalGrav2}
\Box\phi=-\kappa T\,,\quad\text{with}\quad \kappa:= 4\pi G/c^2.
\end{equation}
The next step would be to find the equations of motion
for the world line $z(\tau)$ of a test particle ($\tau$ is
the proper time and dots refer to differentiation with
respect to $\tau$). The obvious first guess,
\begin{equation}
\label{ScalGrav3}
\ddot z^\mu=-\partial^\mu\phi\,,
\end{equation}
is clearly impossible, since it implies the overly restrictive
integrability condition $\dot z^\mu\partial_\mu\phi=0$.
However, this problem can easily be taken care of by replacing 
the right-hand side of (\ref{ScalGrav3}) with its projection 
orthogonal to $\dot z$:
\begin{equation}
\label{eq:ParticleMotion}
\ddot z^\mu=
-\bigl(\eta^{\mu\nu}+{\dot z}^\mu{\dot z}^\nu/c^2\bigr)
\partial_\nu\phi\,.
\end{equation}
This results in three consistent equations of motion for the
three spatial velocity components (the fourth component is,
as always, determined by $\dot z_\mu\dot z^\mu=-c^2$).

It is instructive to relate the naive theory based on
(\ref{eq:ScalGrav2}) and (\ref{eq:ParticleMotion}) to a more
systematic treatment based on modern methods using the action
principle. Since the physical system under consideration consists
of the gravitational field $\Phi$ (its relation to the field
$\phi$ above will become clear soon) and matter. Hence we have three
basic contributions to the total action,
\begin{equation}
\label{eq:TotAction}
S_{\rm tot}=S_{\rm field}+S_{\rm matter}+S_{\rm int}\,,
\end{equation}
where $S_{\rm field}$ is the action of the free gravitational
field and $S_{\rm int}$ that of the interaction between
the gravitational field and matter. If we assume our $\Phi$
to satisfy equation~(\ref{eq:ScalGrav2}), their sum is given
by\footnote{Note that $\Phi$ has the physical dimension of a squared
velocity, $\kappa$ that of length-over-mass. The prefactor
$1/\kappa c^3$ gives (\ref{eq:ActionFieldInt}) the physical
dimension of an action. The overall signs are chosen according to
the general scheme for Lagrangians: kinetic minus potential energy
(cf.\,footnote\,\ref{foot:T-Sign}).}
\begin{equation}
\label{eq:ActionFieldInt}
S_{\rm field}+S_{\rm int}= \frac{-1}{\kappa c^3}\int d^4x\left(
\tfrac{1}{2}\partial_\mu\Phi\partial^\mu\Phi-\kappa\Phi T\right)\,.
\end{equation}
$S_{\rm matter}$ is the action for the matter system which we
only specify in that we assume that the matter consists
of a point particle of rest-mass $m_0$ and a `rest' that
remains unspecified. Hence, $S_{\rm matter}=S_{\rm particle}+
S_{\rm r.o.m}$ (r.o.m = `rest of matter') where
\begin{equation}
\label{eq:ActionPP}
S_{\rm particle}=
-m_0c^2\int d\tau\,.
\end{equation}
The quantity $d\tau=\tfrac{1}{c}\sqrt{-\eta_{\mu\nu}dz^\mu dz^\nu}$ is the
proper time along the worldline of the particle. The energy-momentum
tensor of the particle is given by
\begin{equation}
\label{eq:T-PointParticle}
T^{\mu\nu}(x)=m_0c\,\int d\tau
{\dot z}^\mu(\tau){\dot z}^\nu(\tau)\ \delta^{(4)}(x-z(\tau))\,,
\end{equation}
so that the particle's contribution to the interaction term in
(\ref{eq:ActionFieldInt}) is
\begin{equation}
\label{eq:InteractionPP}
S_{\text{int-particle}}=-m_0\int d\tau\,\Phi(z(\tau))\,.
\end{equation}
Hence the total action can be written in the following form:
\begin{equation}
\label{eq:ScalGrav4}
\begin{split}
S_{\rm tot}=
& - m_0c^2\int d\tau\ \bigl(1+\Phi(z(\tau))/c^2\bigr)\\
& -\frac{1}{\kappa c^3}\int d^4x\ \bigl(\tfrac{1}{2}\partial_\mu\Phi\partial^\mu\Phi
  -\kappa\Phi T_{\text{r.o.m}}\bigr)\\
& +S_{\text{r.o.m}}\,.
\end{split}
\end{equation}

We can now relate this theory to the preceding one.
Recall that, by construction, the field equation for $\Phi$ that
follows from (\ref{eq:ScalGrav4}) is just (\ref{eq:ScalGrav2})
with $\phi$ replaced by $\Phi$. But the analogous statement
is not true for the equation of motion for the point particle.
In fact,  variation of (\ref{eq:ScalGrav4}) with respect to $z(\tau)$
gives (\ref{eq:ParticleMotion}), where
\begin{equation}
\label{eq:PhiphiRelation}
\phi=c^2\ln(1+\Phi/c^2)\,.
\end{equation}
Because of (\ref{eq:ParticleMotion}) it is more natural to call
$\phi$ the gravitational potential. For example, $-\vec\nabla\phi$
is the force on a unit test-mass. Summing up, we may say that a
systematic treatment retains (\ref{eq:ParticleMotion}) but
replaces (\ref{eq:ScalGrav2}) with the same equation in terms of
$\Phi$, whose relation to $\phi$ is (\ref{eq:PhiphiRelation}).
In linear approximation $\phi=\Phi$ and we do get back to the
naive theory.

Note that the action for the point particle with interaction may
be interpreted in various ways. One is to say that the inertial
mass is changed from $m_0$ to  $m \equiv m_0\,e^{\phi/c^2}$
by the interaction with the gravitational field, thereby becoming
spacetime dependent. This was in fact one of Einstein's
concerns:
\begin{quote}
``The law of motion of the mass point in a gravitational field had also
to be adapted to the special theory of relativity. The path was not so
unmistakably marked out here, since the inert mass of a body might
depend on the gravitational potential. In fact, this was to be expected
on account of the principle of the inertia of energy.''
(Einstein, 1977 p.\,135)
\end{quote}
Another interpretation, later (1914) considered by Einstein and 
Fokker (CPAE, Vol.\,4, Doc.\,28), is that the particle moves 
inertially, though not in the Minkowski metric but a conformally 
rescaled metric: 
$\eta_{\mu\nu}\rightarrow g_{\mu\nu}:=e^{2\phi/c^2}\,\eta_{\mu\nu}$. 
This law, which is independent of the particle's rest mass, gives a 
strong hint that ``geometrization'' is a perfect scheme to achieve 
a universal coupling of gravity to matter.

The scalar theory outlined so far clearly satisfies the weak equivalence
principle, according to which all freely falling pointlike
test-masses\footnote{A `test-mass' should have vanishing electric charge,
vanishing intrinsic angular momentum (spin), and vanishing higher
(than zeroth) multipole moments of its mass distribution.} move on the
same world line for given initial data (spacetime point and four velocity).
But this does not imply that the acceleration in a gravitational field is
independent of the center-of-mass motion, such as, e.g.,  an initial horizontal
velocity. To see this, assume the gravitational field is static in some
frame. Then the particle's equation of motion is equivalent to the
following 3-vector equation (a dot now signifies a derivative with
respect to coordinate time $t$)
\begin{equation}
\label{eq:ScalGrav5} \ddot{\vec z}=-\bigl(1-\vert\dot{\vec z}\vert^2/c^2\bigr)\
\vec\nabla\phi,
\end{equation}
which is almost like the Newtonian equation, were it not for
the additional term in parentheses on the right-hand side, which
diminishes the vertical acceleration at high particle velocities.
Although this is a quadratic effect in $v/c$, Einstein considered
this to be a very serious failure of the scalar theory of
gravitation, which made him abandon that track. He wrote:
\begin{quote}
``These investigations, however, led to a result which raised my
strong suspicion. According to classical mechanics, the vertical
acceleration of a body in the vertical gravitational field is
independent of the horizontal component of its velocity.
Hence in such a gravitational field the vertical acceleration
of a mechanical system or of its center of gravity comes out
independently of its internal kinetic energy. But in the theory
I advanced, the acceleration of a falling body was not
independent of its horizontal velocity or the internal energy
of the system. This did not fit with the old experimental
fact that all bodies have the same acceleration in a
gravitational field.''
(Einstein, 1977 pp.\,135--136)
\end{quote}
The dependence of the vertical acceleration on the horizontal
center-of-mass velocity is clearly expressed by (\ref{eq:ScalGrav5}).
However, Einstein's additional claim that there is also a similar
dependence on the internal energy does not survive closer scrutiny.
One might think at first that (\ref{eq:ScalGrav5}) also predicts that,
e.g., the gravitational acceleration of a box filled with gas molecules
is less when heated up, due to the larger velocities of the
gas molecules. But this arguments neglects the walls of the box which gain
in stress due to the rising gas pressure, and according to
(\ref{eq:ScalGrav2}) more stress means less weight. In fact, a
general argument due to Laue (1911) shows that these effects
precisely cancel (for detailed discussion, see Norton, 1993).

We want to draw attention to the remarkable closing section of
Einstein's part of the already mentioned ``Entwurf'' paper, written  
jointly with Grossmann, entitled: ``Can gravity be described by 
a scalar?'' (CPAE, Vol.\,4, Doc.\,13, p.\,321-323)). Via an 
apparently simple Gedanken experiment Einstein implicitly claims 
to show that any scalar theory of gravity, in which the trace of 
the energy-momentum tensor acts as source, necessarily violates 
energy conservation. For the modern field theorist this is a 
surprising statement indeed, for any Poincar\'e invariant theory 
has a conserved Noether charge connected with the symmetry
of time-translations. This general argument was not
available to Einstein (Noether's seminal paper only appeared in 1918),
but it does show that Einstein's argument cannot be taken at
face value. Closer inspection shows that in theories such as the one 
described by (\ref{eq:ScalGrav4}), the conserved energy contains a 
contribution in which the local stresses within a body couple to 
the local gravitational potential. It seems that this contribution 
is not taken into account properly in Einstein's argument. More on 
this will appear elsewhere (Giulini,~2005c).

In any case, arguments of various kinds seem to have triggered a
conceptual phase transition in Einstein's thinking. He now adopted
the strict equivalence principle rather than Lorentz invariance as
his major guiding principle. During his time in Prague, this
led him to consider non-linear modifications of (\ref{eq:ScalGrav0}),
such as the following:
\begin{equation}
\label{eq:ScalGrav6}
\Delta\phi=\kappa\phi\left(
\rho+\frac{(\vec\nabla\phi)^2}{2\kappa\,\phi^2}\right)\,,
\end{equation}
where $\phi$ is now required to approach the value $1$ rather than $0$
at large distances from the source. This equation may be derived from 
the requirement that the self-energy of the gravitational field acts 
as source on a par with the energy density of matter (see Giulini~1997).
However, in Einstein's treatment the field $\phi$ is interpreted as a 
(spatially) variable velocity of light. This put him in opposition to 
contemporaries such as Gunnar Nordstr\"om, Gustav Mie, and Max Abraham 
who still searched for a special relativistic theory of gravity (though
Abraham's theory also contained a variable speed of light).

\subsection{The Poincar\'e invariant approach}
What theory of gravitation would have emerged from
the attempts of Abraham, Nordstr\"om, and Mie? What would
have happened if Einstein had left physics in, say,
1912? Would have GR never have come into being?

We do not think so, but presumably it would have been discovered
much later in a non-geometrical way that is often called the
``flat field approach to gravitation''. In 1939 Fierz \& Pauli
discussed, as an example of their work on higher spin equations,
the field equation for a free massless spin-2 field. These authors
were well aware of the difficulties that arise when a spin-2 field
is coupled to matter. After this initial step, the idea that GR can be
formulated as a consistent, highly non-linear spin-2 theory in flat
spacetime was repeatedly studied. The first published work in this
direction seems to be that by Gupta (1954, 1957) and
Kraichnan (1955, 1956). Fierz also seems to have been thinking about this idea
early on, which much later led to the thesis work
of Wyss~(1965). Other early attempts were made by Thirring,
who was advocating this approach with different emphasis
in various talks and publications (e.g., Thirring, 1961).
Fortunately, Feynman's Caltech lectures on gravitation, which also
emphasize the field theoretic approach, have become available
in book form (Feynman 1995). Weinberg  (1964a, 1964b) also tried to develop a
quantum theory of a self-interacting spin-2 field on flat spacetime.
(We now know that such theories are not renormalizable, and neither are their
supersymmetric extensions.) The theme was taken up later
by Deser~(1970), Wald~(1996), and others.
Quite recently it was shown that one cannot have several, mutually
interacting spin-2 fields (Boulanger \emph{et\,al.}~2001).
This is important for string theory, where one identifies gravity
with a massless spin-2 field.

As already discussed, the simplest possibility of gravitational
theory in flat spacetime is that of a scalar field. 
Since such theories predict no global light deflection,\footnote{ 
See Ehlers \& Rindler (1997) for a discussion of the difference
between local and global light deflection.} Einstein urged 
astronomers in 1913 to measure the light deflection during the 
solar eclipse the following year in the Crimea. Moreover, scalar 
theories predict a retrogression of Mercury's perihelion, which in 
case of the theory described by (\ref{eq:ScalGrav4}) is 1/6 of the 
the size of the advance predicted by GR.\footnote{The same retrogression
is also predicted by Nordstr\"om's ``second'' theory (Nordstr\"om 1913),
whereas Nordstr\"oms ``first'' theory  (Nordstr\"om 1912) predicts 
twice that value (Roseveare 1982, p.\ 153). Of course, this only became 
a problem for these theories after 1915 when GR correctly predicted the 
perihelion advance (CPAE, Vol.\ 6, Doc.\ 24). The earlier 
``Entwurf'' theory predicted an advance 5/12 the size of the GR 
value (CPAE, Vol.\,4, Doc.\,14).}

A spin-1 theory is also not viable. Such a theory would
essentially be given by Maxwell's equations, with one
appropriate sign change in order to make like charges (masses)
attract rather than repel one another. But this leads to a sign change
in the expression for the field's energy, which then becomes
unbounded from below, giving rise to potential
instabilities. The perihelion advance it predicts is
1/6 the Einsteinian value, again in contrast to observation.
 within the parentheses
One is thus led to a spin-2 theory. If one tries the simplest
version by coupling a spin-2 field $h_{\mu\nu}$ linearly to
the energy-momentum tensor of matter, the resulting field
equation is unphysical. Since the free spin-2 theory has
a gauge symmetry, the field equation implies that
$\partial_\nu T^{\mu\nu}=0$, which is unacceptable.
For instance, the motion of a fluid would not be
affected by the gravitational field in that case. Clearly, one has to
include \emph{back-reactions} on matter, which makes the theory
\emph{non-linear}. From the results in the works cited above it follows
that there is only one consistent way of doing this.
The gauge group of the linear theory has to be extended to
the full diffeomorphism group, and the field equations become
equivalent to Einstein's equations for a Lorentz metric
determined by the spin-2 field $h_{\mu\nu}$.

At this point one can re-interpret the theory geometrically.
Thereby the flat metric disappears completely and one arrives
at GR (cf.\ Mittelstaedt, 1970, and references therein).
In summary we can say this: The natural development of the
theory shows that it is possible to eliminate the flat Minkowski
metric, leading to a description in terms of a curved metric
which has a direct physical meaning. The originally postulated
Lorentz invariance turns out to be physically meaningless and
plays no useful role. The flat Minkowski spacetime becomes a
kind of unobservable ether. The conclusion is inevitable that
spacetime is a Lorentzian manifold with a the metric that is a
dynamical field subject to the Einstein field equations.

\section{Einstein's theory of spacetime and gravity}
\subsection{General Remarks}
After some detours, which we cannot describe here, Einstein
arrived at the final form of GR in November 1915.
It is a geometric field-theory par excellence. No non-dynamical
background structures exist, and its equations are invariant
under the largest group possible: the group of spacetime
diffeomorphisms. However, elements of this group do not play the
role of symmetries, as the Lorentz transformations did in
SR, but of gauge transformations. As stressed above,
this means that any two field configurations connected by a
diffeomorphism are empirically indistinguishable and thus
physically identical.

The fundamental field is a Lorentzian (pseudo Riemannian) metric
$g_{\mu\nu}$ on a four-dimensional manifold $M$, obeying a system
of ten non-linear (but quasi-linear) differential equations
($G$ is again Newton's constant):
\begin{equation}
\label{eq:EinsteinEquations}
G_{\mu\nu}= (8\pi G/c^4)\,T_{\mu\nu}\,.
\end{equation}
Here we adopted the signature convention `mostly plus', i.e.
$(-,+,+,+)$, and neglected a possible cosmological term which
we will introduce later. The `Einstein Tensor', $G_{\mu\nu}$,
is a second order differential expression in the metric components
$g_{\mu\nu}$ and directly relates to its
curvature.\footnote{Taking the trace of the Riemannian curvature
Tensor $R^\alpha_{\phantom{\alpha}\mu\beta\nu}$ in $\alpha\beta$
one gets the Ricci tensor $R_{\mu\nu}$. Contracting the Ricci tensor
with $g^{\mu\nu}$ one obtains the Ricci scalar $R$. The Einstein tensor
is now defined  by $G_{\mu\nu}:=R_{\mu\nu}-\tfrac{1}{2}g_{\mu\nu}R$.}
$T_{\mu\nu}$ is the stress-energy tensor of matter, which
generally also involves $g_{\mu\nu}$.

Given a solution $g_{\mu\nu}$, a spinless test particle moves on
geodesics of that metric, which is therefore best compared to the
gravitational \emph{potential}. The idea of a gravitational \emph{field}
is then played by the connection $\Gamma_{\mu\nu}^\lambda$,
which appears in the geodesic equation:
\begin{equation}
\label{eq:GeodesicEquation}
{\ddot x}^\lambda+\Gamma_{\mu\nu}^\lambda {\dot x}^\mu {\dot x}^\nu=0\,,
\end{equation}
and which is determined by $g_{\mu\nu}$ and its first derivatives.
The conceptual difference to other `fields' is that the connection is
not a tensor field on spacetime. This can be seen as a consequence of
the equivalence principle, according to which $\Gamma_{\mu\nu}^\lambda$
vanishes locally in a freely falling frame.

Another difference to other fields is that gravity is not a `force' in
the Newtonian sense. In Newtonian physics, a force is the cause for
deviations from inertial motion. But (\ref{eq:GeodesicEquation})
\emph{defines} inertial motion and the `gravitational field',
$\Gamma_{\mu\nu}^\lambda$, is a structural prerequisite for such a
definition. Again this is a consequence of unifying inertia and gravity.

\subsection{Some current theoretical problems of GR}
\label{sec:CurrentTheorProblems}
The Einstein field equations~(\ref{eq:EinsteinEquations})
are at the core of GR and much research over the last 50 years has gone into their mathematical
analysis. One of the main issues has been
whether the equations admit a well posed initial-value
formulation (Cauchy Problem), as many physical questions are
naturally addressed that way. This turned out to be the case,
albeit in a slightly more complicated fashion due to general
diffeomorphism invariance. Roughly speaking, four of the ten
components of (\ref{eq:EinsteinEquations}) are mere restrictions
on the initial data, so-called ``constraints'', and the remaining
six components are evolution equations. This means that given initial
data for $g_{\mu\nu}$ which satisfy the constraints, Einstein's
equations leave undetermined the evolution of four out of the ten
components of $g_{\mu\nu}$. However, this does not reflect any
lack of physical predictability, but merely the existence of gauge
redundancies corresponding to arbitrary point transformations, which
account for the four arbitrary functions. Such a situation occurs in
any gauge theory. A pedagogical outline is given, e.g., in (Giulini, 2003).

One of the most prominent features of Einstein's equations is
their non-linearity. This means that solutions evolving from regular
initial data may develop singularities in a finite time. As a result of this, not
much is known about the existence of (temporally) global solutions. Given a
singularity-free solution  generated by
some initial data (i.e., a particular spacetime), it is natural to ask whether sufficiently nearby
(in a suitable sense) data still evolve without the formation of
singularities.\footnote{One says that the evolution has no singularities,
if, technically speaking, the maximal Cauchy development is geodesically
complete.} In this case the original solution is called \emph{stable}.
Instabilities are well-known from hydrodynamics, e.g.,  those due to the
formation of shock waves. One may likewise expect gravitating systems
to be generically unstable due to gravitational shock-waves and
gravitational collapse. It may thus be considered a pleasant surprise that stability results have been obtained. Most importantly,
in a veritable tour de force Christodoulou \& Klainerman (1993) were
able to prove the stability of Minkowski space. Earlier,
Friedrich (1986) had already proven the stability of De\,Sitter space
(a solution to the matter-free Einstein equations with positive
cosmological constant). A few more, rather scattered stability
results exist concerning other cosmological models.

The formation of singularities is, to a certain extent, generic in GR 
(see, e.g.,  Hawking \& Ellis 1973). Singularities might give rise to a
true breakdown of predictability if the singularity is not causally
disconnected from the outside world (i.e., from observers not falling 
into the singularity) by the formation of an event horizon. 
The ``cosmic censorship hypothesis'' expresses the expectation that 
under certain reasonable conditions such a breakdown of predictability 
does not occur. This hypothesis is not yet proven. Part of the problem 
is that it is difficult to formalize. See the review by Clarke (1994) 
for a precise formulation and an account of what has been achieved
so far. A lucid and less technical discussion of the fundamental 
concepts is given by Earman (1995). The notion of a
singularity itself is already far from being straightforward (see Geroch, 1968). 
Often the existence of singularities is demonstrated indirectly 
through \emph{reductio ad absurdum} arguments. But this does not 
give any insight into their formation and structure.

Analytical problems concerning the large-scale behavior of
gravitational fields are currently attracting a lot of interest (see,
e.g., Chru\'sciel \& Friedrich, 2004). Specific results on
black-holes will briefly be reviewed in Sec.\,\ref{sec:RelAstro}.

Other analytical problems with more direct relevance for experiments,
such as the ongoing search for gravitational waves, concern the motion
of compact bodies in the strong-field regime. Here one
particularly wishes to understand the phases in which most of the
gravitational radiation is generated. Good candidates for
such generation processes are the close encounters and mergers
of neutron stars and black holes. Analytically, the case of
black holes is simpler, for it can be described by the matter-free
Einstein equations. But it still is a genuine field-theoretic problem,
as point objects do not exist in GR (a feature that to some extent
is mimicked by (\ref{eq:ScalGrav6}); see Giulini, 2003). There is
no known analytical solution to the two-body problem so that a combination of
refined analytical approximation schemes and numerical techniques
becomes essential for evolving initial data. But what are the appropriate
initial data for two black holes in close proximity, from which we can reliably
calculate (numerically) the flux of gravitational waves produced
in the merging process? There are two problems here: First,
numerical methods are used to integrate certain field components in the near-zone,
whereas the mathematical identification of gravitational radiation
is done in the far zone (on ``future null-infinity'', should it exist).
No unambiguous analytical procedure relating the former to the
latter (i.e., in the form of a flux-theorem) has been given.
Strictly speaking,  however, this is exactly what is needed to relate
the integration in the near zone to an actual energy loss of the
system. Secondly, standard data for two black holes, even the most
simple ones describing two non-rotating holes momentarily at
rest, seem to be filled with gravitational radiation up to
spatial infinity already (Valiente-Kroon, 2003). Hence it seems that one
either needs to distinguish between radiation already contained
in the data and radiation produced by the merger---and it is hard to see how that could be done---or
to modify data outside the two-hole region, so that it no longer
contains radiation. It was only understood recently
(Corvino, 2000) that initial data can, in fact, be modified
locally, which is not obvious.\footnote{It is not obvious because the
data have to keep satisfying the constraints, which form an
underdetermined elliptic system of differential equations.
Note that solutions to strictly elliptic systems generally
cannot be modified locally, since they are uniquely determined
by the boundary conditions.} But so far this existence result
has not been backed up by sufficiently concrete methods that
could be employed to change initial data in a physically
controllable way.

For other aspects of current activities in mathematical
relativity, see (Frauendiener \emph{et\,al.}, 2006).

\subsection{Some aspects of the current experimental situation}
During the last few years we have seen tremendous developments in experimental
and observational gravity. These range from weak-field tests using
planets or satellites in earthbound or solar-system orbits, to very
impressive strong-field tests on Galactic binary-systems with compact
objects, like neutron stars. We shall have more to say about the latter
systems in Sec.\,\ref{sec:RelAstro}. Here we make some comments on
the classic weak-field tests.

In Sec.\,\ref{sec:SRTexp} we roughly described how qualitative
statements about a range of aspects of a theory can be made by
parameterizing possible deviations and extracting upper bounds
for their values from observations. Various methods
to do this in GR have been developed to a high degree
of sophistication (see, e.g.,  Will 1993). One of them is the
so-called ``Parameterized Post-Newtonian'' (PPN) formalism, where
one considers finite-parameter families of metrics, $g_{\mu\nu}$,
all of which are within a post-Newtonian approximation scheme. Each
of the parameters may be thought of as probing a specific
deviation from the prediction made by GR. Some correspond
to so-called ``preferred-frame'' and ``preferred-location''
effects, others parameterize possible violations of conservation
of total momentum. If we discard all those, only two
parameters $\beta$ and $\gamma$ remain (the so-called
``Eddington-Robertson parameters'').

Consider the case of a static, spherically symmetric metric:
\begin{equation}
\label{eq:EddingtonParam1}
ds^2=g_{00}(r)\,c^2\,dt^2+g_{ab}(r)\,dx^a\,dx^b\,.
\end{equation}
Its parameterized form contains $\gamma$ and $\beta$, which measure
spatial curvature and ``non linearity'',\footnote{Being ``non linear''
depends on the coordinates used, i.e., is a gauge dependent
statement.} respectively:
\begin{equation}
\label{eq:EddingtonParam2}
\begin{split}
g_{00}(r)&\,=\,-\left[1-2(m/r)+2\beta (m/r)^2+O([m/r]^3)\right]\,, \\
g_{ab}(r)&\,=\,\left[1+2\gamma(m/r)+O([m/r]^2)\right]\,\delta_{ab}\,.\\
\end{split}
\end{equation}
Here $m \equiv GM/c^2$, where $M$ is the central mass, $G$ is Newton's constant,
and $c$ is the velocity of light. The physical dimension of $m$ is that
of length. In GR (without cosmological constant),
the unique static and spherically symmetric solution is the
Schwarzschild solution, which corresponds to the values $\gamma=\beta=1$.

Typical experiments testing the value of $\gamma$ involve the
deflection of light's direction of travel. For (\ref{eq:EddingtonParam2})
the deflection angle comes out to be
\begin{equation}
\label{eq:LightDefl}
\Delta{\theta}=\tfrac{1}{2}(1+\gamma)\cdot
\underbrace{\frac{4m}{d}}_{\text{GR value}}\,,
\end{equation}
where $d$ is the impact parameter (distance of closest approach).
Using Very Long Baseline Interferometry (VLBI ), one finds that
deflections of various astronomical sources lead to the upper bound
$\vert\gamma -1\vert<2\cdot 10^{-4}$.

More accurate tests involve the delay in integrated time of
propagation, when a signal travels between two sites at distances
$r_1$ and $r_2$ from the central body, with closest approach $d$
to the body (the so-called Shapiro time delay). For a round trip
the delay is given by:
\begin{equation}
\label{eq:TimeDelay}
\Delta T=\tfrac{1}{2}(1+\gamma)\cdot
\underbrace{\frac{4m}{c}\cdot
\ln\bigl(4r_1r_2/d^2\bigr)}_{\text{GR value}}\,.
\end{equation}
What is directly observed is not individual delay times, but
their variation in observation-time $t$, as the line of sight
(and hence $d$) comes closer to the central body (the Sun).
The best available data currently available were obtained
with the Cassini spacecraft on its flight to Saturn in June-July 2002. Stable
and coherent two-way radio signals were exchanged and fractional
frequency shifts were observed. This led to the upper bound (Bertotti \emph{et\,al.}, 2003):
\begin{equation}
\label{eq:CassiniData}
\vert\gamma-1\vert<4.4\cdot 10^{-5}\,.
\end{equation}

A typical observable effect which is sensitive to the value of
$\beta$ is the ``anomalous'' advance, or shift, of the
periastron. The advance $\Delta\varphi$ per revolution,
calculated for the metric (\ref{eq:EddingtonParam2}),
is given by
\begin{equation}
\label{eq:PeriastronAdv}
\Delta\varphi=\tfrac{1}{3}(2\gamma-\beta+2)\cdot
\underbrace{\frac{6\pi m}{a(1-\varepsilon^2)}}_{\text{GR value}}\,,
\end{equation}
where $a$ is the radius of the semi-major axis and
$\varepsilon$ is the eccentricity. Applied to solar system planets
(in which case one speaks of the perihelion), this shift is
most pronounced for the innermost planet, Mercury, and quite
accurately known (corresponding to a localization of Mercury
up to 300 meters using radar reflection techniques). In order
to compare observations with (\ref{eq:PeriastronAdv})
one has to take into account other effects contributing to
the perihelion shift. Those originating from perturbations
of other planets\footnote{To these perturbations, Venus, Jupiter,
and Earth make the largest contributions with 52, 29, and 17
percent respectively.} have been known in the 19th century.
It was Einstein's very first triumph with GR to show that
it accounts precisely for the discrepancy (of about 8 percent)
between the observed shift and the shift due to planetary
perturbations.

However, Einstein did not take into account a possible contribution
from the Sun's quadrupole moment. Such a contribution would be significant, if the quadrupole moment, which is measured
by a dimensionless number $J_2$, were at the upper end of the interval
considered plausible for our Sun, which is roughly the interval
$10^{-7}<J_2<10^{-5}$. Ironically, this stirred up considerable
controversy during the 1960s and 70s, with one side arguing that the motion of Mercury's perihelion
refuted rather than confirmed
GR! Essentially the problem in the traditional approach is that
estimations of $J_2$ are made on the basis of the relation between the Sun's oblateness and its
surface angular velocity, a procedure that depends on one's model of the Sun.
In particular, it involves assumptions about its interior state of
differential rotation. Modern results all suggest a
`small' value of $J_2$ of about $2\cdot 10^{-7}$
(e.g., Lydon \& Sofia, 1996). This is confirmed by new methods that
use normal modes of solar oscillations (helioseismology) in order
to get information about the internal structure of the Sun
(e.g., Roxburgh, 2001). See also Pireaux \emph{et\,al.} (2003) for
a comprehensive discussion and many references.

A small value for $J_2$ implies that the contribution
of the Sun's quadrupole moment to the perihelion shift is less
than $10^{-3}$ times the relativistic effect. Combining this value with the results of
radar observations on Mercury and with the upper
bound (\ref{eq:CassiniData}) on $\vert\gamma-1\vert$, one finds the
following upper bound for the parameter $\beta$:
\begin{equation}
\label{eq:MercuryData}
\vert \beta-1\vert < 3\cdot 10^{-3}\,.
\end{equation}
Note that if the orbiting mass is not a test body,
but comparable in mass to the central body, (\ref{eq:PeriastronAdv})
still applies  as long as $m$ is now taken to be the \emph{sum}, $m_1+m_2$,
of the two masses. In this form (\ref{eq:PeriastronAdv}) is applied to
binary-pulsar systems.

Other modern tests are sensitive to the intrinsic angular
momentum (spin) of the central body. In that case the static
metric (\ref{eq:EddingtonParam1}) needs to be generalized to
a stationary one (for constant angular momentum), which differs
from (\ref{eq:EddingtonParam1}-\ref{eq:EddingtonParam2}) by an
off-diagonal contribution. In the simplified version of the
PPN formalism displayed here, this contribution takes the form
\begin{equation}
\label{eq:StationaryMetric}
g_{0a}(r)=\tfrac{1}{2}(1+\gamma)
\cdot\frac{2\,(\vec x\times\vec s)_a}{r^3}\,,
\end{equation}
where $ \vec s \equiv \vec SG/c^3$, $\vec S$ being the (constant)
spin vector. The physical dimension of $\vec s$ is that of
length-squared. It is interesting to note that at this level
of approximation and under the simplifications assumed here,
no new PPN parameter
enters.\footnote{This is because
we excluded preferred-frame and preferred-location effects,
as well as violations of total momentum conservation.
Then the ``electric'' and ``magnetic'' parts of the linearized
gravitational field are related by local Lorentz
invariance. This is just as in electrodynamics, where the
magnetic field produced by a moving charge can be obtained from
the Coulomb field of a charge at rest by a Lorentz transformation.
In particular, as in electrodynamics, the split between
`electric' and `magnetic' parts of the gravitational field is
observer-dependent.}

A  technologically very audacious experiment recently completed
and currently being analysed is Gravity-Probe~B. It consists of an 
earthbound satellite in a
polar\footnote{The orbit is chosen polar in order to avoid
unwanted contributions to the measured effect from the Earth's
quadrupole moment.} orbit approximately $640\,km$ above ground.
The satellite contains four magnetically suspended gyroscopes. According to GR, the Earth's rotation should induce a precession of these gyroscopes
(with respect to a Quasar background).

The metric components (\ref{eq:StationaryMetric}) cause local inertial frames
(here realized by drag-free suspended gyroscopes) to rotate with respect to
asymptotic  frames at large radii (here realized by the quasar
background) at an angular frequency of (to leading order):
\begin{equation}
\label{eq:OmegaGyro}
\vec\Omega_{\text{gyro}}(\vec x)=
\tfrac{1}{2}(\gamma+1)\cdot
\underbrace{c\cdot\frac{3\vec n(\vec n\cdot\vec s)
-\vec s}{r^3}}_{\text{GR value}}\,,
\end{equation}
where $\vec n \equiv \vec x/r$. This is called the Lense-Thirring precession.
In the case of Gravity Probe B, the predicted precession is about $10^{-10}\vec\omega$, where $\vec\omega$ is the Earth's angular
velocity. This amounts to a miniscule precession of $47$ milli-arcseconds
per year! The Gravity-Probe-B experiment is expected to verify this
prediction of GR at the one-percent level.\footnote{Another
prediction of GR is the geodetic (or de\,Sitter-) precession,
which is a consequence of spatial curvature and hence directly
sensitive to $\gamma$. In the present situation it
is about 160 times larger than the Lense-Thirring precession.
If completed successfully, Gravity-Probe-B should therefore
measure $\gamma$ with an accuracy of $3\cdot 10^{-5}$, thereby
slightly improving on the accuracy reached by Cassini.} Its conceptual
importance is that it directly measures the gravitational effects
caused by mass-\emph{currents}, sometimes referred to as
``dragging effects''. That part of the gravitational field which
is  generated by mass-currents is often called the
``gravitomagnetic field''. It is needed for theoretical consistency,
just as the magnetic field is needed in electrodynamics
(for a lucid discussion, see Nordtvedt, 1988).

Another dragging effect of spinning central bodies is the precession
of orbital planes around it. This has been recently verified for the
earthbound system of two LAGEOS satellites (designed for other purposes), though only  at the 10\% level (Ciufolini \& Pavlis, 2004).

More accurate though indirect measurements of dragging effects
exist for neutron-star binary systems. These are due to spin-orbit
and (much more pronounced) orbit-orbit couplings. According to GR, a spinning
companion gives a contribution to the periastron shift of
\begin{equation}
\label{eq:OmegaPeriastron}
\vec\Omega_{\text{periastron}}(\vec x)=
-\,2c\cdot\frac{3\vec n(\vec n\cdot\vec s)-\vec s}%
{a^3(1-\varepsilon^2)}\,.
\end{equation}
Here $a$ is the semi-major axis and $\varepsilon$ is the orbital eccentricity.
This means that if spin and orbital angular momentum form an
acute angle, the periastron shifts due to (\ref{eq:OmegaPeriastron}) and due to (\ref{eq:PeriastronAdv})
(where $m$ now corresponds to the sum of masses) will be in opposite directions. For
binary pulsars spin-orbit effects have been calculated by Damour
\& Sch\"afer (1988).

In a binary system with comparable masses, the two components also
move with comparable velocities in the center-of-mass frame. In that case gravitomagnetic
fields of both components,
contribute to their mutual periastron advance (orbit-orbit
coupling). This effect is generally much bigger than that due to
spin. For example, the Hulse-Taylor pulsar shows a total periastron
advance of $4.2^\circ$ per year, which is the sum of about $10^\circ$ per
year from the gravitoelectric and about $-6^\circ$ per year
resulting from dragging due to the orbital motion of each
companion in the center-of-mass frame (see Nordtvedt, 1988).

\subsection{Early history of gauge and Kaluza-Klein theories}
\label{sec:GaugeKKTh}
The history of gauge theories begins with GR, which can be regarded
as a non-Abelian gauge theory of a special type. To a large extent
the other gauge theories gradually emerged, in a slow and complicated process,
from GR. Their common geometrical structure---best
expressed in terms of connections of fiber bundles---is now widely
recognized.

\subsubsection*{Weyl's papers on the gauge principle}
It all began with H. Weyl (1918) who made the first attempt to extend
GR in order to describe gravitation and electromagnetism within a
unifying geometrical framework. This brilliant proposal contains the
germ of all mathematical aspects of non-Abelian gauge theory. The
word `gauge' (german: `Eich') transformation appeared for the first
time in a subsequent paper on this theory (Weyl 1919, p.\ 114; cf.\ CPAE 8, Doc.\ 661, note 5),
but in the everyday meaning of change of length
or change of calibration.

Einstein admired Weyl's theory as ``a coup of genius of the first
rate'' (CPAE, Vol.\,8, Doc.\,498), but immediately realized
that it was physically untenable.
After a long discussion Weyl finally admitted that his attempt was a
failure as a physical theory (for discussion, see Straumann,
1987.) It paved the way, however, for the correct understanding of
gauge invariance. After the advent of quantum theory, Weyl himself reinterpreted his original theory
 in a magisterial paper (Weyl 1929).
This reinterpretation had
actually been suggested before by London (1927). Fock (1926),
Klein (1926), and others arrived at the principle of gauge invariance
in the framework of wave mechanics along  completely different
lines.\footnote{For details see the survey by Jackson and Okun, 2001, which
also discusses the 19th-century roots of gauge invariance.} It was Weyl,
however,  who emphasized the role of gauge invariance as a
\emph{constructive principle} from which electromagnetism can be
derived. This point of view became very fruitful for our present
understanding of fundamental interactions.

Weyl's papers have repeatedly been discussed in detail
(see O'Raifeartaigh \& Straumann, 2000). Weyl's reinterpretation was connected to his
incorporation of Dirac's theory into GR,  an important
contribution in and of itself. This in turn was related to Einstein's recent
unified theory, which invoked a distant parallelism
with torsion. Wigner (1929) and others had noticed a connection
between this theory and the spin theory of the electron.
Weyl did not care for this and wanted to dispense with teleparallelism.
This he achieved with the help of local tetrads (\textit{Vierbeine}),
a technique that had been used extensively before by Cartan. Preparing the ground with a
general-relativistic formulation of spinor theory,
Weyl begins the final section of his 1929 paper with:
\begin{quote}
``We come now to the critical part of the theory. In my opinion
the origin and necessity for the electromagnetic fields is the
following. The components $\psi_1,\psi_2$ [of the two-component
spinor field] are, in fact, not uniquely determined by the tetrad
but only to the extent that they can still be multiplied by an
arbitrary ``gauge factor'' $e^{i\lambda}$. The transformation of
the $\psi$ induced by a rotation of the tetrad is determined only
up to such a factor. In SR one must regard this gauge factor as a
constant because we have only a single point-independent tetrad.
Not so in GR; every point has its own tetrad and hence its own
arbitrary gauge-factor; because by the removal of the rigid
connection between tetrads at different points the gauge-factor
becomes an arbitrary function of position.''
(Weyl, 1968, Vol.\,III, Doc.\,85, p.\,263)
\end{quote}
In this way Weyl arrived at the gauge principle in its modern
form. As he emphasized: ``from the arbitrariness of the gauge factor
in $\psi$ appears the necessity to introduce the electromagnetic
potential ''(Weyl, 1968, Vol.\,III, Doc.\,85, p.\,263).

\subsubsection*{The early work of Kaluza and Klein}
Early in 1919 Einstein received a paper by Theodor Kaluza, a young
mathematician ({\it Privatdozent}) and consummate linguist in
K\"{o}nigsberg. Inspired by the work of Weyl the year before,
Kaluza proposed another geometrical unification of gravitation and
electromagnetism by extending spacetime to a five-dimensional
pseudo-Riemannian manifold. Einstein reacted very positively.
On April 21, 1919 he wrote to Kaluza: ``The idea of achieving
[a unified theory] by means of a five-dimensional cylinder
world never dawned on me (...). At first glance I like your
idea enormously''
(CPAE, Vol.\,9, Doc.\,26). A few weeks later he added: ``the formal unity of your theory is
startling'' (CPAE, Vol.\,9, Doc.\,35). The
fourth of the five letters of Einstein to Kaluza (CPAE, Vol.\,9,
Doc.\,40), however, makes it understandable why Einstein, despite his initial enthusiasm, delayed the publication
of Kaluza's work for almost two years. In this letter Einstein raised a serious objection. What worried Einstein was the  apparently huge
influence of the scalar field on the electron in the dimensional
reduction of the five-dimensional geodesic equation. Einstein expressed
his hope that Kaluza would find a way out. But Einstein's ``serious
difficulty'' (``ernsthafte Schwierigkeit'') remained, as Kaluza (1921) acknowledged in his published paper.

A few years later, shortly after the discovery of the
Schr\"{o}dinger equation, Oskar Klein improved and extended Kaluza's
treatment, and revealed an interesting geometrical interpretation of
gauge transformations (Klein 1926a, 1926b). Applying the formalism
of quantum mechanics to the five-dimensional geodesic, and assuming
periodicity in the extra dimension, he also suggested that ``the
atomicity of the electric charge may be interpreted as a quantum
law'' (Klein 1926b, p.\,516). The extension of the extra dimension
turned out to be comparable to the Planck length. As Klein writes:
\begin{quote}
``The small value of this length together with the periodicity
of the fifth dimension may perhaps be taken as a support of the
theory of Kaluza in the sense that they may explain the non-appearance
of the fifth dimension in ordinary experiments as the result
of averaging over the fifth dimension.''
(Klein, 1926b, p.\,516)
\end{quote}
For further discussion of this early work on higher-dimensional
unification, see, e.g.,  O'Raifeartaigh \& Straumann (2000).

GR also played a crucial role in Pauli's discovery of non-Abelian
gauge theories. (See Pauli's letters to Pais and Yang in Pauli
1985-99, Vol.\,4).
He arrived at all basic equations through dimensional reduction
of a generalization of Kaluza-Klein theory, in which the internal
space becomes a two-sphere. (For a description in modern language,
see O'Raifeartaigh and Straumann 2000).

In contrast, in the work of Yang and Mills (1954) GR played no role.
In an interview in 1991 Yang recalled:
\begin{quote}
``It happened that one semester [around 1970] I was teaching
GR, and I noticed that the formula in gauge theory for the field
strength and the formula in Riemannian geometry for the Riemann tensor
are not just similar -- they are, in fact, the same if one makes the
right identification of symbols! It is hard to describe the thrill I
felt at understanding this point.''
(Zhang, 1993, p.\,17)
\end{quote}
The developments after 1958 consisted in the gradual recognition
that---contrary to phenomenological appearances---Yang-Mills gauge
theory could describe weak and strong interactions.
Since this history is recounted in numerous textbooks, there is no
need for us to dwell on it.

\subsection{Relativistic astrophysics}
\label{sec:RelAstro}
By 1915 it was known through the work of W. Adams on the binary
system of Sirius that Sirius\,B has an enormous average density of
about $10^6~g/cm^3$. The existence of such compact stars constituted
one of the major puzzles of astrophysics until the quantum
statistical theory of the electron gas was worked out. On August 26,
1926, a paper by Dirac (1926) containing the Fermi-Dirac
distribution was communicated  to the Royal Society by R.H.~Fowler.
On November 3 of the same year, Fowler presented his own work to the
Royal Society (Fowler,\, 1926a), in which he systematically worked
out the quantum statistics of identical particles and, in the
process, developed the well-known Darwin-Fowler method. Shortly
thereafter, on December 10, Fowler (1926b) communicated the Royal
Astronomical Society a new paper with the title ``Dense Matter''. In
this work he showed that the electron gas in Sirius\,B is almost
completely degenerate in the sense of the new Fermi-Dirac
statistics, realizing that ``the black-dwarf is best likened to a
single gigantic molecule in its lowest quantum state'' (Fowler
1926b, p.\,122), and he developed the non-relativistic theory of
white dwarfs. The Fowler theory of white dwarfs is equivalent to the
Thomas-Fermi theory, in which a white dwarf is considered as a big
``atom'' with about $10^{57}$ electrons. For white dwarfs the
(semi-classical) Thomas-Fermi approximation is perfectly
justified.\footnote{The paper by Thomas (1926) was presented at the
Cambridge Philosophical Society on November 6, 1926. (Fermi's work
was independent, but about one year later.) Fowler communicated his
important paper on the non-relativistic theory of white dwarfs about
one month later. One wonders who first noticed the close connection
of the two approaches.}

It is remarkable that the quantum statistics of identical particles,
satisfying Pauli's exclusion principle, found their first application 
in astrophysics. We recall that this principle implies that a 
sufficiently dense gas of such particles builds up a ``zero-point'' 
or ``Fermi'' pressure, depending only on its density and not its 
temperature. If the Fermi pressure dominates the pressure of the 
gas one calls it ``degenerate''. In the ``non-relativistic'' regime, 
where the kinetic energy of each particle is proportional to the 
square of its momentum, the Fermi pressure is proportional to 
the density of the gas raised to the power of $5/3$. 
However, in the ``ultrarelativistic'' regime, the kinetic energy 
becomes directly proportional to the modulus of the momentum, 
as seen from  (\ref{eq:MomentumMinkSquare2}). As a result, the 
Fermi pressure turns proportional to the density raised to the 
power of only $4/3$, a distinctly slower increase. In this 
sense SR has a destabilizing effect, which leads to a finite limiting 
mass for white dwarfs, given roughly by the ratio 
$M_{Pl}^3/m_N^2$. (Here $M_{Pl}$ denotes the Planck mass and $m_N$ 
the nucleon mass.) The existence of such a limiting mass is thus 
an immediate consequence of SR and the Pauli principle.
All this was recognized independently by several people 
(I.~Frenkel, E.~Stoner, S.~Chandrasekhar and L.D.~Landau)
soon after the initial step was taken by Fowler.

In 1934, Chandrasekhar derived the exact relation between mass and 
radius for completely degenerate configurations. He concluded his 
paper with the following statement:
\begin{quote}
``The life-history of a star of small mass must be
essentially different from the life-history of a star of large mass.
For a star of small mass, the natural white-dwarf stage is an initial
step towards complete extinction. A star of large mass cannot pass
into the white-dwarf stage and one is left speculating on other
possibilities.''
(Chandrasekhar, 1934, p.\,77)
\end{quote}
The delayed acceptance of the discovery by the 19-year-old
Chandrasekhar, that quantum theory plus SR imply the existence
of a limiting mass for white dwarfs is one of the more bizarre stories
of the history of astrophysics. The following reaction of Landau is
particularly astonishing:
\begin{quote}
``For $M>1.5M_\odot$ there exists in the whole quantum theory
no cause preventing the system from collapsing to a point. As in
reality such masses exist quietly as stars and do not show any such
ridiculous tendencies, we must conclude that all stars heavier than
$1.5M_\odot$ certainly possess regions in which the laws of quantum 
mechanics (and therefore of quantum statistics) are violated.''
(Israel, 1987, p.\,215)
\end{quote}
Still reeling from the quantum revolution a few years earlier, some
physicists already expected a new revolution in the domain of relativistic
quantum theory.

It is worth mentioning that  Lieb and  Yau (1987) have shown that
Chandrasekhar's theory can be obtained as a limit of a quantum-mechanical
description in terms of a semi-relativistic Hamiltonian.

Soon after the discovery of the neutron, Baade and Zwicky, in a 
remarkable pair of papers (Baade \& Zwicky, 1934a, 1934b),
developed  the idea of a neutron star and made the prescient suggestion 
that such stars would be formed in supernova explosions:
\begin{quote}
``With all reserve we advance the view that supernovae
represent the transitions of an ordinary star into a
\emph{neutron star}, consisting mainly of neutrons. Such a
star may possess a very small radius and an extremely high
density.''
(Baade and Zwicky, 1934b, p.\,263)
\end{quote}
The first  calculations for models of neutron stars in GR were performed by Oppenheimer \& G.~Volkoff (1939). In their pioneering work, they used
the equation of state of a completely degenerate ideal neutron gas. In those early days the effects of strong interactions
could not be estimated. Theoretical interest in neutron
stars soon dwindled, since no relevant observations existed. For two
decades, Zwicky was one of the few who took seriously the probable
role of neutron stars as final states of massive stars. Interest
in the subject was reawakened in the late 1950s and early 1960s. When pulsars were discovered in 1967, especially when
a pulsar with a short period of 0.033\,$s$ was found in the Crab Nebula,
it became clear that neutron stars can be formed in type II
supernova events through the collapse of the stellar core to
nuclear densities. Since then the physics and astronomy of neutron
stars has become one of the major fields of relativistic astrophysics.

\begin{figure}[htb]
\centering\epsfig{figure=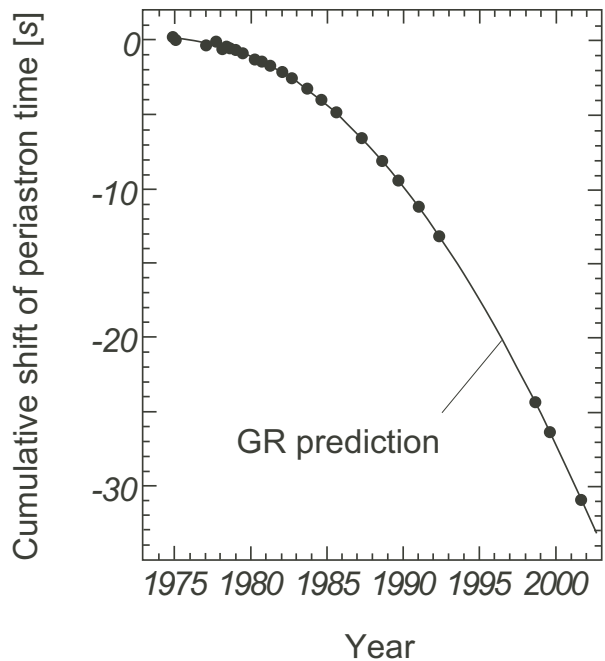,width=0.6\linewidth}
\caption{\small Cumulative shift of periastron time for the
Hulse-Taylor pulsar according to observation (dots) and theory
(solid line). The theoretical prediction takes into
account the energy loss due to the emission of gravitational radiation.
\label{fig:HT-period}}
\end{figure}

Systems in close proximity containing two neutron stars (binary and double pulsars)
have led to the most remarkable tests of GR. One of them is the
celebrated Hulse-Taylor pulsar PSR\,1913+16 that we have already mentioned,
which gave rise to the first indirect evidence of
gravitational waves. The measured long-term decrease of its orbital
period agrees perfectly with the energy loss  due to
the radiation of gravitational waves predicted by GR (see Fig.\,\ref{fig:HT-period}).

Another very interesting system is J0737-3039, which in October 2003 was shown to consist
of two pulsars with pulse periods
of about $23$ milliseconds and $2.7$ seconds, respectively, and an orbital
period of $2.4$ hours, and with an extremely high periastron advance
of almost $17$ degrees per year. This is about four times larger
than that of the Hulse-Taylor pulsar. Fortuitously, the system's
orbital orientation relative to our line of sight is almost exactly
edge-on, which means that measurements of Shapiro time-delay
of pulse periods of one component in the gravitational potential
of the other can be performed with high precision. Presently the
measurements of Shapiro delay verify GR predictions at the 0.1\%
level (Kramer\,\emph{et\,al.} 2005). More results on this exciting
system are expected in the near future.

The theory of black holes belongs to the most beautiful applications
of GR. The structure of stationary black holes was
completely clarified during a relatively short period of time. When
matter disappears behind a horizon, an exterior observer sees almost
nothing of its properties. One can no longer say, for example, how
many baryons formed the black hole. A huge amount of information
thus seems lost. The mass and angular momentum completely determine
the external field, which is known analytically (Kerr solution).
This led Wheeler to say that ``a black hole has no hair'' 
(Wheeler, 1971, p.\,191-192). The preceding statement is now 
known as the \emph{no-hair-theorem}.

The proof of this theorem is an outstanding contribution to mathematical
physics, and was completed in the span of only a few years
by various authors (Israel, Carter, Hawking, and Robinson).
A decisive first  step was taken by Israel (1967), who was able to
show that a \emph{static} black-hole solution of Einstein's vacuum
equation has to be \emph{spherically symmetric} and, therefore, agree
with the Schwarzschild solution. In a second paper Israel extended
this result to black-hole solutions of the coupled Einstein-Maxwell
system. The Reissner-Nordstr\"om $2$-parameter family, it turned out,
exhausts the class of static so-called electrovac black holes. It was then conjectured
by Israel, Penrose and Wheeler that in the stationary case the
electrovac black holes should all be given by the $3$-parameter
Kerr-Newman family. After a number of steps, supplied by various authors,
this conjecture could finally be proven. See (Heusler, 1996) for
a comprehensive account of black-hole uniqueness results.

The evidence for black holes in some X-ray binary systems and for
super-massive black holes in galactic centers is still indirect,
but has become overwhelming during the past few years. However,
there is  little evidence so far that these collapsed objects are
described by the Kerr metric.

Until a few years ago the best one could say about the evidence
for super-massive black holes in the center of some galaxies, was
that it was compelling if dynamical studies and observations of
active galactic nuclei were taken together. In the meantime the
situation has improved radically. The beautiful work of Genzel
and his coworkers has established a dark-mass concentration of about
$3\times10^{6}M_\odot$ near the center of the Milky Way with an extension
of less than $17$ light hours (see, e.g., Ott
\emph{et\,al.}, 2003).
If this were a cluster of low mass stars or neutron stars, its
central density would exceed $10^{17}M_\odot/pc^3$ and would not
survive for more than a few $10^5$ years. The least exotic
interpretation of this enormous dark mass concentration is that
it is a black hole. But this is not the only possibility.
Although alternative interpretations are highly implausible, they
illustrate the point that dynamical studies alone cannot give an
incontrovertible proof of the existence of black holes. Ideally, one
would like to show that some black hole candidate actually has an
\emph{event horizon}. There have been various attempts in this
direction, but presumably only gravitational wave astronomy will
reveal the essential properties of black holes.

Most astrophysicists do not worry about possible remaining doubts.
The evidence for (super-massive) black holes has become so overwhelming
that the burden of proof is now on the hard-core
skeptics.

\subsection{Relativistic cosmology}
\label{sec:RelCosm}
In 1917 Einstein applied GR for the first time to cosmology, and
found the first cosmological solution of a consistent theory of
gravity (CPAE, Vol.\,6, Doc.\,43). In spite of its drawbacks this bold step
can be regarded as the beginning of modern cosmology. It is still
interesting to read this paper about which Einstein says:
``I shall conduct the reader over the road that I have myself
travelled, rather a rough and winding road, because otherwise I
cannot hope that he will take much interest in the result at the
end of the journey.'' In a letter to Ehrenfest of February 4,
1917 (CPAE, Vol.\,8, Doc.\,294), Einstein wrote about his attempt:
``I have again perpetrated something relating to the theory
of gravitation that might endanger me of being committed to a madhouse."\footnote{``Ich habe wieder
etwas verbrochen in der Gravitationstheorie, was mich ein wenig in Gefahr bringt, in ein
Tollhaus interniert zu werden.''}

In his attempt Einstein assumed---and this was completely novel---that
space is globally \emph{closed}. This was because he believed at the 
time that this was the only way to satisfy what he later 
(CPAE, Vol.\,7, Doc.\,4) named \emph{ Mach's principle}, the 
requirement that the metric field be determined uniquely by the 
energy-momentum tensor. In these early years, and for quite some 
time, Mach's ideas on the origin of inertia played an important 
role in Einstein's thinking (for a discussion, see, e.g., Janssen, 
2005). This may even be the
primary reason that he turned to cosmology so soon after the 
completion of GR. Einstein was convinced that isolated
masses cannot impose a structure on space at infinity. His intention
was to eliminate all vestiges of absolute space. It is for such reasons
that he postulated a universe that is spatially finite and
closed, a universe in which no boundary conditions are needed.
Einstein was already thinking about the problem regarding the
choice of boundary conditions at infinity in spring 1916.
In a letter to Michele Besso from May 14, 1916 (CPAE, Vol.\,8, Doc.\,219)
he mentions the possibility of the world being finite.
A few month later he developed these ideas in correspondence with 
Willem de Sitter.

Einstein assumed that the Universe was not only closed but also \emph{static}.
This was not unreasonable at the time, because the relative velocities
of the stars as observed were small.\footnote{Recall that astronomers only
learned later that spiral nebulae are independent star systems
outside the Milky Way. This was definitively established when Hubble found  in
1924 that there were Cepheid variables in Andromeda
as well as in other galaxies. Five years later he announced the
recession of galaxies.}

These two assumptions, however, were incompatible with Einstein's
original field equations. For this reason, Einstein added the
famous $\Lambda$-term, which is compatible with the principles of
GR, in particular with the energy-momentum law
$\nabla_\nu T^{\mu\nu}=0$ for matter. The modified field
equations are (compare (\ref{eq:EinsteinEquations}))
\begin{equation}
G_{\mu\nu} =(8\pi G/c^4) T_{\mu\nu} - \Lambda g_{\mu\nu}\,.
\end{equation}
The cosmological term is, in four dimensions, the only possible
addition to the field equations if no higher than second order
derivatives of the metric are allowed (Lovelock's theorem; see
Lovelock (1971)).
This remarkable uniqueness is one of the most attractive features
of general relativity. (In higher dimensions additional terms
satisfying this requirement are allowed.)

For the static Einstein universe the field equations imply the
two relations
\begin{equation}
(4\pi G/c^2)\rho = \frac{1}{a^2} = \Lambda\,,
\end{equation}
where $\rho$ is the mass density of the dust filled universe
(zero pressure) and $a$ is the radius of curvature. (In passing
we remark that the Einstein universe is the only static
dust solution; one does not have to assume isotropy or homogeneity.
Its instability was demonstrated by Lema\^{\i}tre in 1927.)
Einstein was very pleased by this direct connection between the
mass density and geometry, because he thought that this was in
accord with Mach's philosophy.

Einstein concludes with the following sentences:
\begin{quote}
``In order to arrive at this consistent view, we
admittedly had to introduce an extension of the field equations
of gravitation which is not justified by our actual knowledge
of gravitation. It has to be emphasized, however, that a
positive curvature of space is given by our results, even if
the supplementary term is not introduced. That term
is necessary only for the purpose of making possible a quasi-static
distribution of matter, as required by the fact of the small
velocities of the stars.''
(CPAE, Vol.\,6, Doc,\, 43, p.\,551)
\end{quote}

In a letter to De\,Sitter of March 12, 1917
(CPAE, Vol.\,8, Doc.\,311), Einstein emphasized that his model was 
intended primarily to settle the question ``whether the basic idea 
of relativity can be followed through its completion, or whether 
it leads to contradictions''. He added that is was an entirely 
different matter whether the model corresponds to reality.

Only later did Einstein come to realize that Mach's philosophy is
predicated on an antiquated ontology that seeks to reduce the metric
field to an epiphenomenon of matter. It became increasingly clear to
him that the metric field has an independent existence
(corresponding to physical degrees of freedom), and his enthusiasm
for  Mach's principle gradually evaporated. In a letter to Pirani he
wrote in 1954: ``As a matter of fact, one should no longer speak of
Mach's principle at all." (Pais, 1982, Sec.\,15e).\footnote{``Von
dem Machschen Prinzip sollte man eigentlich \"uberhaupt nicht mehr
sprechen.''} The absolute existence of the spacetime continuum,
independent of any matter, is a remnant in GR of Newton's absolute
space and time. For a modern and comprehensive discussion of various
aspects of Mach's principle and their status in GR, see (Barbour \& Pfister, 1995).

\subsubsection*{From static to expanding world models}
It must have come as quite a shock to Einstein, that within days of 
receiving a letter in which Einstein described his cosmological model 
(CPAE, Vol.\,8, Doc.\,311), De Sitter had found a completely different 
cosmological model---also allowed by the new field equations with 
cosmological term---that was \emph{anti-Machian} in that it contained 
no matter whatsoever (CPAE, Vol.\,8, Doc.\,312; De Sitter 1917a). 
For this reason, Einstein tried to discard it on various grounds 
(more on this below). Einstein and De Sitter mostly discussed 
De Sitter's solution in its so-called static form (CPAE, Vol.\,8, 
Doc.\,355; De Sitter 1917b):
\begin{equation}
\label{eq:DeSitterMetric}
ds^2 = -\Bigl[ 1 - \left(\frac{r}{R}\right)^2 \Bigr] c^2dt^2
+ \frac{dr^2}{1-(\frac{r}{R})^2}
+ r^2(d\vartheta^2
+ \sin^2\vartheta d\varphi^2 )\,.
\end{equation}
The spatial metric is that of a three-sphere of radius $R$,
determined by $\Lambda = 3/R^2$. The model had one very
interesting property: For light sources moving along static world
lines there is a gravitational redshift, which became known as the
\emph{De Sitter effect}. This was thought to have some bearing on the
redshift results obtained by Slipher. Because the fundamental (static)
worldlines in this model are not geodesics, a freely- falling particle
released by any static observer  accelerates
away from such an observer, generating local velocity (Doppler) redshifts
corresponding to \emph{peculiar velocities}. In his famous book,
``The Mathematical Theory of Relativity'', Eddington wrote about
this:
\begin{quote}
``De Sitter's theory gives a double explanation for this motion
of recession; first, there is the general tendency to scatter (...);
second, there is a general displacement of spectral lines to the red
in distant objects owing to the slowing down of atomic vibrations
(...), which would be erroneously interpreted as a motion of recession.''
(Eddington, 1924, p.\,161)
\end{quote}
We do not want to enter into all the confusion over the De Sitter
universe (see, e.g., Vol.\,8, pp.\,351--357, the editorial note,
``The Einstein-De Sitter-Weyl-Klein Debate''). One source of confusion was the apparent singularity at
$r=R=(3/\Lambda)^{1/2}$. This was thoroughly misunderstood
at first  even by Einstein and Weyl. In the end,
Einstein had to acknowledge that De Sitter's solution is fully regular
and matter-free and thus indeed a counterexample to Mach's principle.
But he still discarded the solution as physically irrelevant because
it is not globally static. This is clearly expressed in a letter from
Weyl to Klein, dated February 7, 1919 (quoted in CPAE, Vol.\,8,  Doc.\,567, note 3),
after Weyl had discussed the issue during a visit of Einstein to Z\"urich.
An important discussion of the redshift of
galaxies in De Sitter's model by Weyl in 1923 should be mentioned.
Weyl (1923) introduced an expanding version of the De Sitter model.
For \emph{small} distances his result reduced to what later became
known as the Hubble law.\footnote{We recall that the de Sitter model
has many different interpretations, depending on the class of
fundamental observers that is singled out. This point was first
stressed by Lanczos (1922).} Independently of Weyl,
Lanczos (1922) also introduced  a non-stationary interpretation of
De Sitter's solution in the form of a Friedmann spacetime with
positive spatial curvature. In a subsequent paper he
also derived the redshift for the non-stationary interpretation
(Lanczos 1923).

Until about 1930 almost everybody `knew' that the universe was
static, notwithstanding two fundamental papers by Friedmann (1922, 1924)
 and  independent work by Lema\^{\i}tre (1927).\footnote{For discussion of Lema\^{\i}tre's work,
 see Eisenstaedt, 1993.} These path-breaking
papers were largely ignored. The history of this early
period has---as is often the case---been distorted by some widely
read documents.
Einstein too accepted the idea of an expanding universe only much
later. After Friedmann's first paper, he published a brief note
claiming to have found an error in Friedmann's work; when it was
pointed out to him that it was his error, Einstein published a
retraction of his comment, with a sentence that (fortunately for
him) was deleted before publication: ``[Friedmann's paper] while
mathematically correct is of no physical significance'' (Stachel
2002, p.\,469). In comments to Lema\^{\i}tre during the Solvay
meeting in 1927, Einstein again rejected the expanding universe
solutions as physically unacceptable. According to Lema\^{\i}tre,
Einstein told him: ``Vos calculs sont corrects, mais votre physique
est abominable'' (Sch\"{u}cking, 1993). On the other hand, we
found in the archive of the ETH many years ago a postcard of
Einstein to Weyl from 1923, related to Weyl's reinterpretation of De
Sitter's solution, with the following interesting sentence: ``If
there is no quasi-static world, then away with the cosmological
term.'' This goes to show once again that history is not as simple
as it is often being portrayed.

It is also not well-known that Hubble interpreted his famous results
on the redshift of the radiation emitted by distant `nebulae' in
the framework of the De Sitter model. The general attitude is well
illustrated by the following remark of Eddington at a Royal Society
meeting in January 1930: ``One puzzling question is why there should 
be only two solutions. I suppose the trouble is that people look for 
static solutions'' (Eddington, 1930, p.\,850). Lema\^{\i}tre, 
who had been for a short time a post-doctoral student
of Eddington's, read this remark in a report to the meeting published
in \emph{Observatory}, and wrote to Eddington alerting him to his 1927 paper.
Eddington had seen that paper, but had completely forgotten about it.
But now he was greatly impressed and praised Lema\^{\i}tre's work
in a letter to \emph{Nature}. He also arranged for a translation which
appeared in Monthly Notices of the Royal Astronomical Society (Lema\^{\i}tre 1931).

Lema\^{\i}tre's successful explanation of Hubble's discovery finally
changed the viewpoint of the majority of workers in the field.
At this point Einstein (1931) \emph{rejected the cosmological term 
as superfluous and no longer justified}. At the end of the paper 
he made some remarks about the age
problem which was quite severe without the $\Lambda$-term, since
Hubble's value for the Hubble parameter at the time was about seven times
too large. Einstein, however, was not too worried and suggested two
ways out. First, he pointed out that the matter distribution is in reality
inhomogeneous and that the approximate treatment may be illusionary.
Secondly, he cautioned against   large
extrapolations in time in astronomy.

Einstein repeated his new viewpoint much later (Einstein 1945), and
it was adopted by many other influential workers, e.g., by Pauli
(1958, supplementary note\,19). Whether Einstein really considered 
the introduction of the
$\Lambda$-term as ``the biggest blunder of his life'' as related by
Gamov (1970, p.\,44) appears doubtful to us. In his published work
and extant letters such a strong statement is nowhere to be found.
Einstein discarded the cosmological term merely for reasons of
simplicity. For a minority of cosmologists this was not sufficient
reason. Paraphrasing Rabi\footnote{When Rabi heard at a conference
the first time that the muon had been discovered, his reaction was:
``who has ordered it?" (see, e.g., Feynman, 1985,\, p. 165).} one could
ask: `who ordered it away'?

Einstein's paper (1931) was his last one on cosmology. In hindsight
it is somewhat puzzling that after he saw the first paper by
Friedmann he did not realize that his own static model was unstable.

\subsubsection*{Vacuum energy and gravity}
So much for  the classical discussion of the $\Lambda$-term; we do,
however, want to add a few remarks  the $\Lambda$-problem in the
context of quantum theory, where the problem becomes very serious
indeed. Since quantum physicists were facing so many other problems,
it need not surprise us that in the early years they did not worry
about this particular one. An exception was Pauli, who wondered in
the early 1920s whether the zero-point energy of the radiation field
could have significant gravitational effects. He estimated the
influence of the zero-point energy of the electromagnetic radiation
field---cut off at the classical electron radius---on the radius of
the universe, and came to the conclusion that the ``could not even
reach to the moon'' (for more on this, see Straumann 2003a). Pauli's
only published remark on his considerations can be found in his {\it
Handbuch} article  on quantum mechanics, in the section on the
quantization of the radiation field, where he says: ``Also, as is
obvious from experience, the [zero-point energy] does not produce
any gravitational field'' (Pauli, 1933, p.\,250).

For decades nobody else seems to have worried about contributions
of quantum fluctuations to the cosmological constant, although
physicists learned after Dirac's hole theory that the vacuum state
in quantum field theory is not empty but has interesting
physical properties. As far as we know, the first one to come back to
possible contributions of the vacuum energy density to the
cosmological constant was Zel'dovich. He discussed this issue in
two papers (Zel'dovich, 1967, 1968) during the third renaissance period of the
$\Lambda$-term, but before the advent of spontaneously-broken
gauge theories. He pointed out that, even if one assumes in a completely ad-hoc
fashion that the
zero-point contributions to the vacuum energy density are
exactly cancelled by a bare term, there still remain higher-order
effects. In particular, \emph{gravitational} interactions between
the particles in the vacuum fluctuations are expected on
dimensional grounds to lead to a gravitational self-energy
density of order $G\mu^6$, where $\mu$ is some cut-off scale.
Even for $\mu$ as low as 1 GeV, this is about
9 orders of magnitude larger than the observational bound.

This strongly suggests that there is something profound that we do not
seem to understand at all, certainly not in quantum field theory
(nor, at least so far, in string theory).  We are unable to calculate
the vacuum energy density in quantum field theories, such as the
Standard Model of particle physics. But we can attempt to make
what appear to be reasonable order-of-magnitude estimates for the
various contributions. All expectations  are
\emph{in dramatic conflict with the facts} (see, e.g., Straumann, 2003b).
Trying  to arrange the cosmological constant to be zero is unnatural
in a technical sense. It is like enforcing a particle to be massless,
by fine-tuning the parameters of the theory when there is no
symmetry principle implying a vanishing mass. The vacuum
energy density is unprotected from large quantum corrections.
This problem is particularly severe in field theories with
spontaneous symmetry breaking. In such models there are usually
several possible vacuum states with different energy densities.
Furthermore, the energy density is determined by what is called
the effective potential, and this is a {\it dynamical} object.
Nobody can see any reason why the vacuum of the Standard Model we
ended up as the universe cooled, has---by the
standards of particle physics---an almost vanishing energy density. 
Most likely, we shall only find a satisfactory answer once we have 
a theory that successfully combines the concepts and laws of
GR with those of quantum theory.

For a number of years now, cosmology has been going through a fruitful
and exciting period. Some of the developments are clearly
of general interest, well beyond the fields of astrophysics and cosmology.
Lack of space prevents us from even indicating the most 
important issues of current interest.

\subsection*{Acknowledgements}
We sincerely thank Thibault Damour for making valuable
suggestions for improvements on an earlier draft version
and G\"unther Rasche for carefully reading the manuscript.

\end{document}